\documentclass[onecolumn]{aastex631}

\pdfoutput=1

\shorttitle{Forbush Decreases at MESSENGER}
\shortauthors{E.E.Davies et al.}
\graphicspath{{./}{figures/}}
\begin{document}

\title{Characterizing ICME-related Forbush Decreases at Mercury using MESSENGER Observations: Identification of a One or Two-Step Structure}

\author[0000-0001-9992-8471]{Emma E. Davies}
\affiliation{Institute for the Study of Earth, Ocean, and Space, University of New Hampshire, Durham, New Hampshire, USA}
\correspondingauthor{Emma E. Davies}
\email{emma.davies@unh.edu}

\author[0000-0002-9276-9487]{Réka M. Winslow}
\affiliation{Institute for the Study of Earth, Ocean, and Space, University of New Hampshire, Durham, New Hampshire, USA}

\author[0000-0002-7696-6667]{David J. Lawrence}
\affiliation{The Johns Hopkins University Applied Physics Laboratory, Laurel, Maryland, USA}

\begin{abstract}

The large-scale magnetic structure of interplanetary coronal mass ejections (ICMEs) has been shown to cause decreases in the galactic cosmic ray (GCR) flux measured in situ by spacecraft, known as Forbush decreases (Fds). We use measurements of the GCR count rate obtained by MESSENGER during its orbital phase around Mercury to identify such Fds related to the passage of ICMEs and characterize their structure. Of the 42 ICMEs with corresponding high-quality GCR data, 79\% are associated with a Fd. Thus a total of 33 ICME-related Fds were identified, 24 of which (73\%) have a two-step structure. We use a superposed epoch analysis to build an average Fd profile at MESSENGER and find that despite the variability of individual events, a two-step structure is produced and is directly linked with the magnetic boundaries of the ICME. By using results from previous studies at Earth and Mars, we also address whether two-step Fds are more commonly observed closer to the Sun; we found that although likely, this is not conclusive when comparing to the wide range of results of previous studies conducted at Earth. Finally, we find that the percentage decrease in GCR flux of the Fd is greater at MESSENGER on average than at Earth and Mars, decreasing with increasing heliocentric distance. The relationship between the percentage decrease and maximum hourly decrease is also in agreement with previous studies, and follows trends relating to the expansion of ICMEs as they propagate through the heliosphere.

\end{abstract}

\keywords{Solar coronal mass ejections(310) --- Forbush decrease(546) --- Galactic cosmic rays(567) --- Heliosphere(711)}

\section{Introduction} \label{sec:intro}

Forbush decreases \citep[Fds;][]{forbush1937effects} are identified as temporary decreases in the galactic cosmic ray (GCR) flux. Such decreases are associated with solar wind transients such as interplanetary coronal mass ejections (ICMEs), corotating interaction regions (CIRs), and the discontinuities that they drive. Generally, the passage of such transients causes a rapid decrease in the GCR flux, followed by a slower recovery phase typically lasting many days \citep[e.g.][]{cane2000coronal}.

ICMEs are the interplanetary counterparts of coronal mass ejections (CMEs) which are large-scale structures of plasma and magnetic field that erupt from the solar atmosphere and subsequently propagate through the heliosphere. ICMEs often have a clear structure comprising a shock or discontinuity, a turbulent sheath region, and a region of magnetic ejecta (ME) \citep[e.g.][]{zurbuchen2006situ}. In some ICMEs, the ME may exhibit a magnetic flux rope structure, with a smooth rotation of the magnetic field vector, low plasma $\beta$ and a low proton temperature; such ME are known as magnetic clouds \citep[MCs;][]{burlaga1981magnetic}.

The modulation of GCRs by ICMEs presents an indirect approach by which to study the evolution of ICMEs as they propagate. The modulation by different substructures of the ICME have been linked to the profiles of the Fds they drive: one-step profiles display a steady decrease in GCR flux from onset to the minimum of the Fd, whereas two-step profiles display two different decreases between onset and minimum, either of different gradients or separated by a leveling of the GCR flux \citep[e.g.][]{cane1994cosmic}. For a two-step Fd structure \citep[e.g.][]{cane1994cosmic, cane1996cosmic}, the sheath (rather than only the shock front) is the main driver of the initial decrease in GCR flux \citep[see discussion in][]{vonforstner2020comparing}, and the ME has been shown to contribute to the second decrease in GCR flux \citep[e.g.][]{lockwood1991forbush, cane1993cosmic}. If just one of the ICME substructures encounters the spacecraft, this model suggests that a one-step Fd structure would occur. It is a topic of debate as to which Fd structure is more common: \citet{cane1996cosmic} studied 30 years of neutron monitor data between 1964 and 1994, and classified 70\% of Fds driven by ICMEs as two-step Fds. However, more recent studies have found that the two-step structure of a Fd is not always clearly observed and may be less commonplace than previously suggested \citep{jordan2011revisiting}.

The mechanisms by which the different ICME substructures modulate GCRs can be differentiated \citep[e.g.][]{barnden1973largescale, barnden1973forbush} and have been implemented in different Fd models to further explore such processes. The turbulence of the sheath region is thought to increase diffusion and thus affect GCR transport across the region. Previous models of Fds have therefore focused on solving the diffusion-convection equation \citep[e.g.][]{bland1976simple, nishida1982numerical}, and more recently, \citet{luo2017numerical} devised a more complex model using a three-dimensional (3D) diffusion barrier to construct a time-dependent, cosmic ray transport model including diffusion, drift, solar wind convection, and adiabatic cooling. The strong magnetic field strength and closed field line geometry of the ME has been proposed to act as a shield to GCR diffusion, and thereby further reduce the measured GCR flux \citep{krittinatham2009drift}. This mechanism is modeled by \citet{cane1995response} and \citet{richardson2011galactic}, solving the diffusion equation for particles entering the ME through perpendicular diffusion. Similarly, the more recent ForbMod model proposed by \citet{dumbovic2018analytical, dumbovic2020evolution} also considers the perpendicular diffusion of particles that fill the closed magnetic structure of the ME, but goes further by also incorporating flux rope expansion to create a more realistic Fd profile caused by the ME. 
To better understand the role different substructures play in driving the Fd, studies have performed superposed epoch analyses (SEA) to characterize the average profile of ICME and Fd properties at ACE and at the Earth using neutron monitor data, respectively \citep{masias2016superposed, janvier2021twostep}. \citet{masias2016superposed} investigated how the velocity of the ICMEs affected the average profile of the Fd, finding that the percentage decrease in GCR flux and the recovery period were larger for faster ICMEs. At all velocity ranges, minima of the average GCR flux profiles were found to be located towards the end of the sheath region, however, for faster ICMEs, second minima were also observed within the ME. This result is consistent with the findings of \citet{belov2015galactic}, for which a local minimum of GCR flux was observed for faster ICMEs with magnetic field strengths $>$18~nT ($>$20~nT for those ICMEs with MCs). \citet{janvier2021twostep} investigated the relationship between ICME structure and Fd profiles, comparing SEA profiles of Fds driven by ICMEs with and without sheath regions, finding that those without sheath regions drive Fds that have a symmetrical profile with the minimum GCR flux located closer to the center of the ICME. By comparing ICMEs with and without sheath regions of similar speed profiles, \citet{janvier2021twostep} also found that the intensity of the magnetic field strength is the necessary condition to drive a Fd, rather than fluctuations of the magnetic field. Overall, many ICME parameters may contribute to and correlate with the decrease in GCR flux affecting the Fd structure, including the plasma speed, the magnetic field strength, and the fluctuations of the magnetic field \citep[e.g.][]{dumbovic2011cosmic, kumar2014interplanetary}. 

Fds have been studied most extensively at Earth, predominantly using neutron monitor data as mentioned above. More recently, Fds have also been cataloged at Mars, using both surface measurements from the Radiation Assessment Detector \citep[RAD;][]{hassler2012rad_msl} instrument on the Mars Science Laboratory \citep[MSL;][]{grotzinger2012msl} mission and spacecraft measurements obtained from the SEP instrument \citep[][]{larson2015sep_maven} onboard the Mars Atmosphere and Volatile EvolutioN \citep[MAVEN;][]{jakosky2015maven} mission \citep{guo2018marsfds}. \citet{vonforstner2020comparing} compared the properties of Fds driven by ICMEs between Earth and Mars, finding that Fds show a strong correlation between their minimum GCR flux and the maximum hourly decrease in GCR flux but with different proportionality factors at the different heliocentric distances, attributed to the increase in size of ICMEs as they propagate. Other studies that investigate the evolution of ICME driven Fd properties beyond 1~AU include \citet{witasse2017interplanetary}, which investigated the signatures caused by a single ICME observed at Mars, comet 67P, and Saturn. Comparing observations, \citet{witasse2017interplanetary} found that the profile of the Fd changed with increasing heliocentric distance, with the steepest decrease in GCR flux towards the lowest minimum observed at the heliocentric distance closest to the Sun (in this case at Mars).  

Most studies of Fds in the inner heliosphere have utilized data obtained from the Helios mission, covering a heliocentric distance range of 0.3--1~AU. \citet{cane1997helios} used Helios data to investigate the association between decreases in cosmic ray flux and ICME ejecta and found that 88\% of decreases in flux were associated with ejecta. \citet{blanco2013energetic} also used Helios data to study 19 Fds associated with magnetic clouds, and observed relationships between both the depth of the GCR flux within the Fd and magnetic cloud velocity and magnetic field strength. 

It is also possible to study Fds in the inner heliosphere using the data from the Neutron Spectrometer \citep[NS;][]{goldsten2007mesns} instrument onboard the MESSENGER spacecraft, which was in orbit around Mercury from 2011 to 2015 and covered a heliocentric distance range of 0.31--0.47~AU: \citet{winslow2018window} identified an ICME with corresponding Fds observed in conjunction by MESSENGER at Mercury, LRO at the Moon, the SOPO Neutron Monitor at Earth, and MSL at Mars. The signatures indicated a two-step Fd at MESSENGER, which is not observed clearly at the Moon/Earth nor Mars. From this study, the authors found that the Fd size decreases exponentially with heliocentric distance, and hypothesized that two-step Fds are more common closer to the Sun. 

In this study, we build upon \citet{winslow2018window} by compiling a comprehensive database of all ICME-related Fds observed by MESSENGER during its orbital phase around Mercury when NS data were available. We build an average Fd profile at MESSENGER using a superposed epoch analysis, and attempt to relate the two-step Fd structure to the ICME substructure. By using results from previous studies at Earth and Mars, we also aim to address whether two-step Fds are more commonly observed closer to the Sun, and how Fd profiles and properties change with increasing heliocentric distance. Section \ref{sec:gcr_dataset} presents the MESSENGER data used in this study and Section \ref{sec:fd_identification} details how Fds were identified. A summary of Fd properties and relationships between different measures is presented in Section \ref{sec:properties}. Similarly to \citet{masias2016superposed} and \citet{janvier2021twostep}, we implement the SEA technique to produce average profiles of Fds observed at MESSENGER, and discuss the implications of the results of the study in Section \ref{sec:discussion}. 

\section{The MESSENGER GCR Dataset} \label{sec:gcr_dataset}

\begin{figure*}[t!]
\centering
\includegraphics[width = \textwidth]{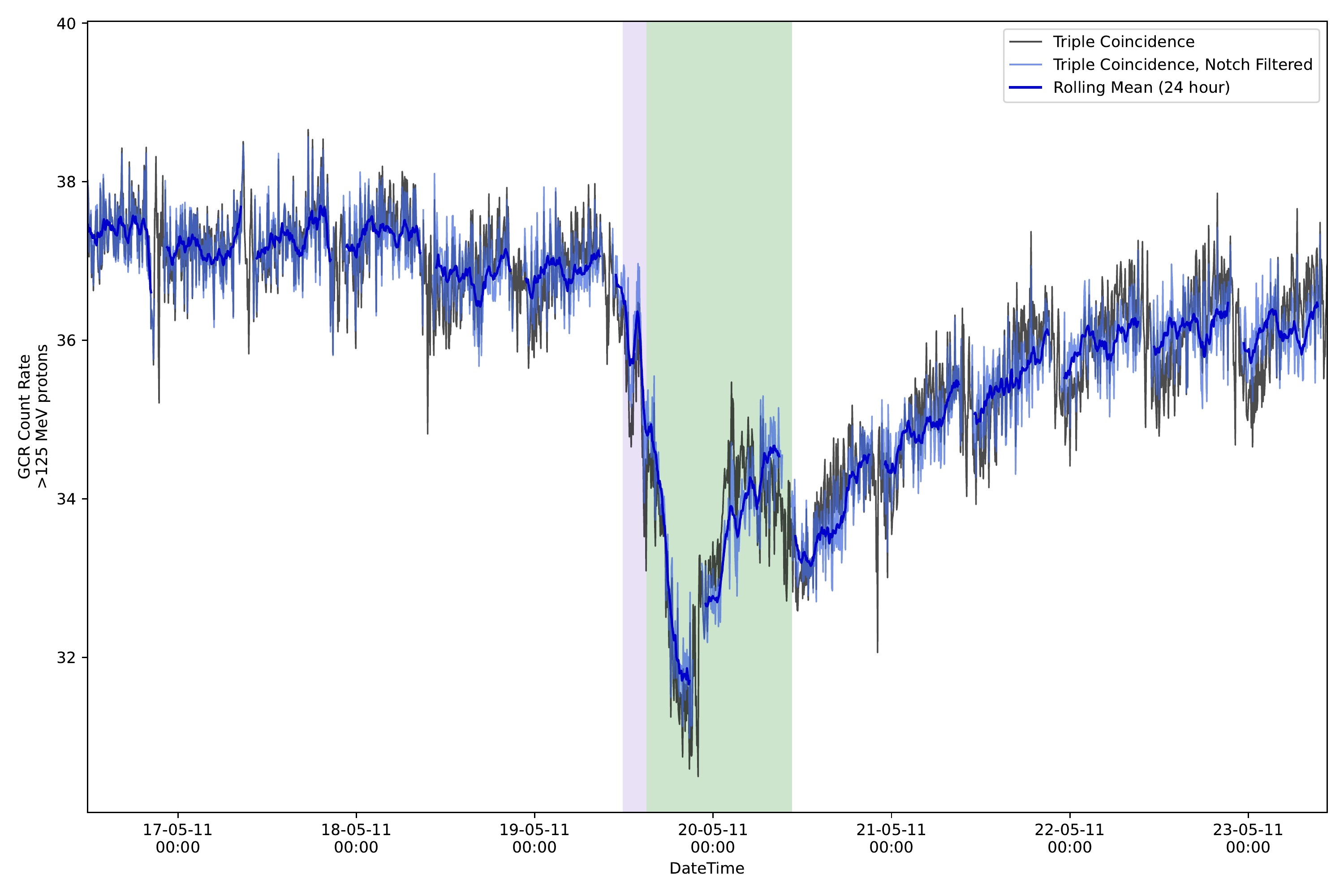}
\caption{The first Forbush decrease (Fd) identified in the MESSENGER GCR dataset during May 2011. The dark grey line presents the original triple-coincidence GCR count rate measured by the Neutron Spectometer (NS), demonstrating the effect of the variation in altitude during each orbital period (12~hours). The altitude filtered ($>$3000~km) and notch filtered data is shown by the light blue line, where the notch filter attempts to remove the remaining altitudinal variation in GCR count rate. The final dark blue line overlaid is the calculated 24 hour rolling mean of the filtered data, used to aid in defining key times associated with the Fd. The shaded areas in the background represent the ICME sheath (purple) and magnetic ejecta (green) regions identified in the magnetic data at MESSENGER.} 
\label{fig:gcr_data}
\end{figure*}

In this study, we use data obtained from the Neutron Spectrometer \citep[NS;][]{goldsten2007mesns} on board NASA’s MErcury Surface, Space ENvironment, GEochemistry, and Ranging (MESSENGER) spacecraft as a proxy for the galactic cosmic ray (GCR) count rate. The NS was capable of detecting neutrons of energies between 0.4~eV and 10~MeV, energetic particles i.e. protons and alpha particles between 15 ~MeV/nucleon and $>$125~MeV/nucleon, and electrons between 1~MeV and $>$30~MeV \citep{lawrence2014detection, lawrence2016galactic}. The NS was designed to measure neutrons at the surface of Mercury for planetary composition measurements. However, the instrument is also sensitive to GCR protons greater than 125~MeV. These charged particle measurements are used to correct for non-composition variations in the neutron data \citep{lawrence2004gamma, maurice2004reduction, lawrence2016galactic}. Charged particles that deposit energy in all three of the NS scintillators (particles that are triple-coincident) are the highest-energy particles measured by the instrument. From instrument modeling, this triple-coincidence count rate measures the integral flux of protons and alpha particles $>$125 MeV/nucleon and was therefore shown to be a proxy for the GCR flux at MESSENGER. We use the 5-minute resolution triple-coincidence GCR flux count rate to identify Forbush decreases at MESSENGER. 

Between March 2011 and April 2015, MESSENGER was in orbit around Mercury. The orbital period of MESSENGER was initially 12~hours, which decreased to 8~hours from April 2012 onwards \citep{lawrence2015comprehensive}. During each orbit, the altitude of the spacecraft with respect to the surface of Mercury varied introducing a periodic variation of the GCR count rate where values were strongly affected below altitudes of only a few thousand kilometers. Figure \ref{fig:gcr_data} presents the GCR data surrounding a Forbush decrease identified in May 2011. The dark grey line presents the original GCR triple-coincidence count rate for which the altitude variation is clearly observed. Similarly to \citet{winslow2018window}, we filter the data to remove values that correspond to altitudes less than 3000~km in our study. To account for the remaining variation with altitude, we apply a notch filter set to the frequency of the orbital period; notch filters are designed to attenuate a specific frequency component of a signal, whilst preserving frequency components both higher and lower than the frequency which is to be removed. In this case, a notch filter has been applied to attenuate the frequency component of the signal that is related to the orbital period. The notch filtered data for altitudes greater than 3000~km is given by the light blue line in Figure \ref{fig:gcr_data}.

\section{Identification of ICME-related Forbush Decreases in the MESSENGER GCR Dataset} \label{sec:fd_identification}

\begin{figure*}
\gridline{\fig{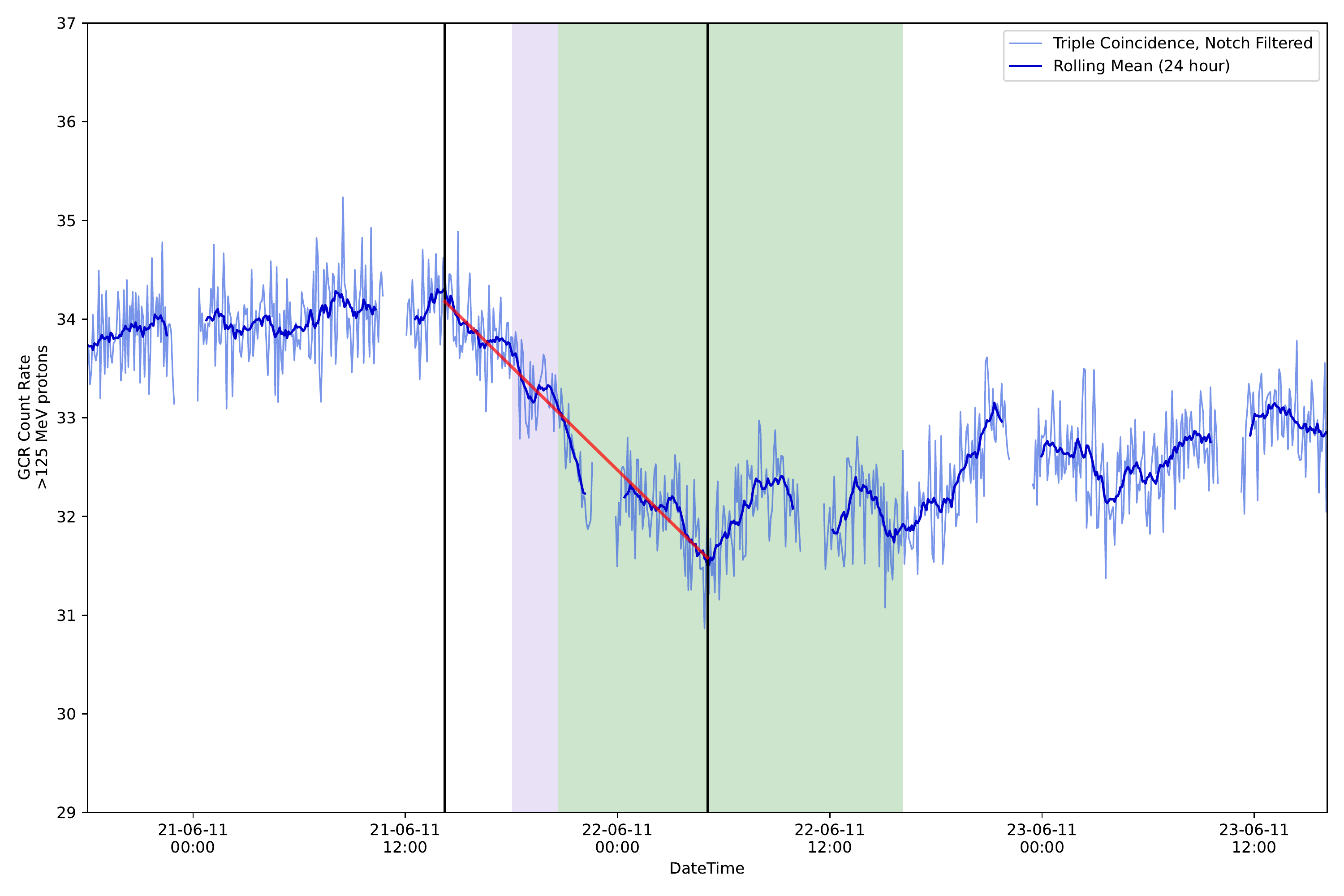}{0.5\textwidth}{(a) One-step: Fd onset 2011-06-21 14:13:43.}
          \fig{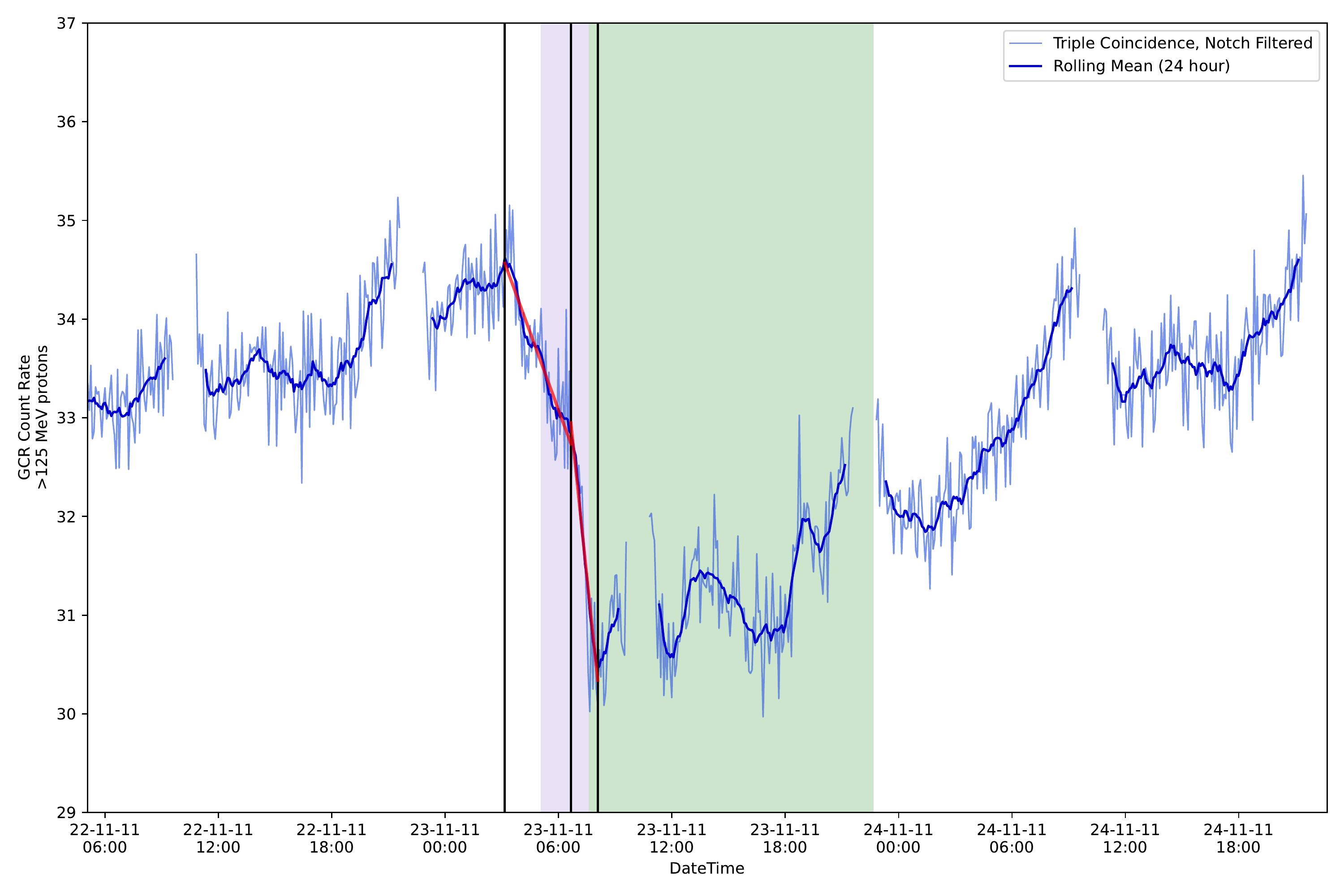}{0.5\textwidth}{(b) Two-step: Fd onset 2011-11-23 03:09:14.}
          }
\caption{Two examples of Forbush decreases identified in the MESSENGER GCR dataset. Similarly to Figure \ref{fig:gcr_data}, the ICME sheath and ME region are shaded in purple and green, respectively. The vertical black lines mark the onset, second drop (Figure \ref{fig:two_examples}b) and minimum times in chronological order. The results of the linear regression fitting are overlaid in red. Figure \ref{fig:two_examples}a presents an example of a one-step Fd, where the red line fits well to the data between onset and minimum. Figure \ref{fig:two_examples}b presents an example of a two-step Fd, where two red line fits have clearly different gradients between onset and minimum.}
\label{fig:two_examples}
\end{figure*}

In this study, we focus on identifying Fds associated with the passage of ICMEs. We therefore utilize the catalogs of \citet{winslow2015interplanetary, winslow2017magnetosphere} which identified ICMEs in the magnetic data \citep[MAG;][]{anderson2007magnetometer} measured at MESSENGER as a starting point to search for decreases in the GCR count rate. We plot 10~day periods of the notch and altitude-filtered GCR data centered on the ICME start time, where the sheath region (defined as the region between the ICME start time and ME start time) is shaded in purple and the ME region (between the ME start and ICME end times) is shaded in green, as demonstrated in Figure \ref{fig:gcr_data}. We use the more generic term ME throughout this paper, as the ICMEs identified at MESSENGER cannot be confirmed as MCs due to the lack of plasma observations. We note any clear decrease in GCR count rate $>$3\% in comparison to the mean preceding background value that occurred within a few hours of the ICME start time as an ICME-related Fd.

Of the 69 ICMEs listed in the catalogs of \citet{winslow2015interplanetary, winslow2017magnetosphere}, 42 had corresponding GCR flux data of good quality in which to identify potential decreases. We identified a total of 33 ICME-related Fds, therefore 79\% of ICMEs that also had GCR data were associated with a Fd. For each Fd identified, we use the calculated 24-hour rolling mean to aid in defining the onset time and the time of the minimum of the Fd. Both the ICME magnetic boundaries and the Fd boundaries are given in our database published in machine-readable format; a sample of these events are presented in Table \ref{tab:database}.

We analyze in detail the Fd structure in order to distinguish between one and two-step Fds. For 73\% (24/33) of the Fds identified, we observe more than one different slope or drop between the onset time and minimum of the Fd: a possible two-step structure or more complex structure (where more than two-steps are possible). For these cases, Table \ref{tab:database} also includes the time of the second drop. To investigate the structure of the Fds further, we perform a linear regression to fit the data between the onset time and the minimum, or two separate linear fits between onset and the second drop time, and between the second drop time and minimum for those with a likely two-step or more complex structure. 

Such examples are presented in Figure \ref{fig:two_examples}, where Figure \ref{fig:two_examples}a displays an example of a one-step Fd and Figure \ref{fig:two_examples}b displays an example of a likely two-step Fd. Similarly to Figure \ref{fig:gcr_data}, the ICME sheath and ME region are shaded in purple and green, respectively. The vertical black lines mark the onset, second drop (where present) and minimum times in chronological order. The results of the linear regression fitting are overlaid in red. For Figure \ref{fig:two_examples}a, the single linear regression line fits well to most of the data, although there is a data gap during which the spacecraft was $<$3000~km in altitude which may miss some of the possible structure. For Figure \ref{fig:two_examples}b, the two linear regression lines fit well to the data, with two clearly different slope gradients between onset and minimum.

Figures \ref{fig:appendix1} and \ref{fig:appendix2} present all the Fds identified in this study in a similar format to Figure \ref{fig:two_examples}, with the equations of each slope listed. The Fds presented demonstrate the variability of individual events and are clearly linked to the ICME substructures that modulate the GCR flux. For example, for many events, we observe a clear initial decrease associated with the ICME sheath region, followed by a plateau before the second decrease commences, either close to the end of the ICME sheath region, or close to the start of the ICME ME region. This leveling of the GCR count rate between steps has previously been identified in Fd profiles at Earth \citep{cane1994cosmic, wibberenz1997twostep, jordan2011revisiting}. 

For each Fd, we note the number of steps e.g. 1, 2, or complex (C), in the database. This is determined by the quality of both the observations themselves and the fitting of the linear regression to the observations. We therefore also give a quality rating between 1 (highest) and 4 (lowest) for each event listed in the database. The quality ratings indicate the confidence in the number of steps identified within the Fd, taking into account how well the linear regression fits to the data, missing data, and whether missing data could obscure the number of steps we define. A quality rating of 1 therefore indicates a Fd with a clear structure, high quality fits, and no data gaps, in comparison to a quality rating of 4, which indicates a low confidence in the number of steps identified due to large data gaps obscuring the structure and poor fits. We define any event with missing data as only able to achieve a maximum quality rating of 2. However, events with a quality rating of 2 still indicate high quality fits and confidence in the number of steps identified, whereas a quality rating of 3 indicates large data gaps and a lower confidence in number of steps identified. 

\begin{splitdeluxetable*}{lllccBlllcccBlcc}
\tablecaption{Sample of the Forbush Decreases identified.}
\label{tab:database}
\tablewidth{0pt}
\tablehead{
\colhead{ICME Start} & \colhead{ME Start} & \colhead{ICME End} & \colhead{Heliocentric Distance} & \colhead{Max B} & \colhead{FD Onset} & \colhead{FD Second Drop} & \colhead{FD Nadir} & \colhead{Drop Duration} & \colhead{Drop \%} & \colhead{m$_{max}$} & \colhead{Step Number} & \colhead{Step Quality} & \colhead{1st:2nd Slope Ratio} \\
%\cline{9-12} \cline{1-6}
\colhead{} & \colhead{} & \colhead{} & \colhead{[AU]} & \colhead{[nT]} & \colhead{} & \colhead{} & \colhead{} & \colhead{[hrs]} & \colhead{} & \colhead{} & \colhead{} & \colhead{} & \colhead{} 
}
\decimals
\startdata
2011-05-19 11:50:02 & 2011-05-19 15:00:07 & 2011-05-20 10:36:28 & 0.41 & 78.7 & 2011-05-19 11:04:27 & 2011-05-19 15:23:47 & 2011-05-19 19:48:47 & 8.74 & 14.0 & 9.3 & C & 1 & 0.50\\
2011-06-21 18:01:47 & 2011-06-21 20:39:02 & 2011-06-22 16:07:08 & 0.33 & 77.2 & 2011-06-21 14:13:43 & \nodata & 2011-06-22 05:04:49 & 14.85 & 8.2 & 6.3 & 1 & 4 & \nodata \\	
2011-10-15 08:26:36 & 2011-10-15 11:13:45 & 2011-10-16 06:43:13 & 0.46 & 89.9 & 2011-10-15 08:16:26 & 2011-10-15 10:36:26 & 2011-10-15 12:11:26 & 3.92 & 16.6 & 12.4 & 2 & 1 & 0.39\\
2011-11-04 15:09:11 & 2011-11-04 17:22:58 & 2011-11-05 15:40:40 & 0.44 & 76.5 & 2011-11-04 15:09:35 & 2011-11-04 18:49:35 & 2011-11-05 02:04:43 & 10.92 & 28.2 & 19.1 & C & 2 & 0.23\\
2011-11-14 02:06:58 & 2011-11-14 08:27:45 & 2011-11-14 22:59:12 & 0.40 & 53.2 & 2011-11-14 01:59:26 & 2011-11-14 07:19:26 & 2011-11-14 16:19:28 & 14.33 & 6.9 & 3.4 & 2 & 3 & 0.16\\
\enddata
\tablecomments{For each Forbush decrease (Fd) identified, we list the associated ICME boundary times i.e. the ICME start, magnetic ejecta (ME) start, and the ICME end times in ISO standard. The heliocentric at which the ICME was observed is given, as well as the maximum magnetic field strength observed within the ICME. The Fd boundary times i.e. onset, second drop (if observed), and nadir times are then listed, alongside the duration between onset and nadir (drop duration), the percentage decrease in GCR count rate between onset and nadir (drop \%), and the maximum hourly percentage decrease are listed (m$_{max}$). Finally, the number of steps observed (1, 2, or C for complex), the quality rating (1--4, from highest to lowest) and the ratio between the 1st and 2nd slope (if present) are given. Table \ref{tab:database} is published in its entirety in machine-readable format. A sample is presented here for guidance regarding its form and content.}
\end{splitdeluxetable*}

\section{Summary of Fd Properties} \label{sec:properties}

For each Fd, the database (Table \ref{tab:database}) also includes the duration between onset time and minimum (drop duration) and the percentage drop in GCR count rate between onset and minimum (drop percentage. We calculate the drop percentage using the difference between values of the 24-hour rolling mean at the defined onset and minimum times, and normalizing by the value at onset. We define both the drop duration and drop percentage between onset and minimum for consistent comparison across all events identified, whether they are noted as a one-, two-step, or complex profile. Similarly, we also calculate the maximum hourly percentage decrease in GCR count for each event. For profiles with more than one step, the ratio between the 1st and 2nd slope gradients is given. Finally, following the ICME magnetic boundaries, we also list the heliocentric distance at which the shock front of the ICME was observed and the maximum magnetic field strength measured within the ICME.

\begin{figure*}
\gridline{\fig{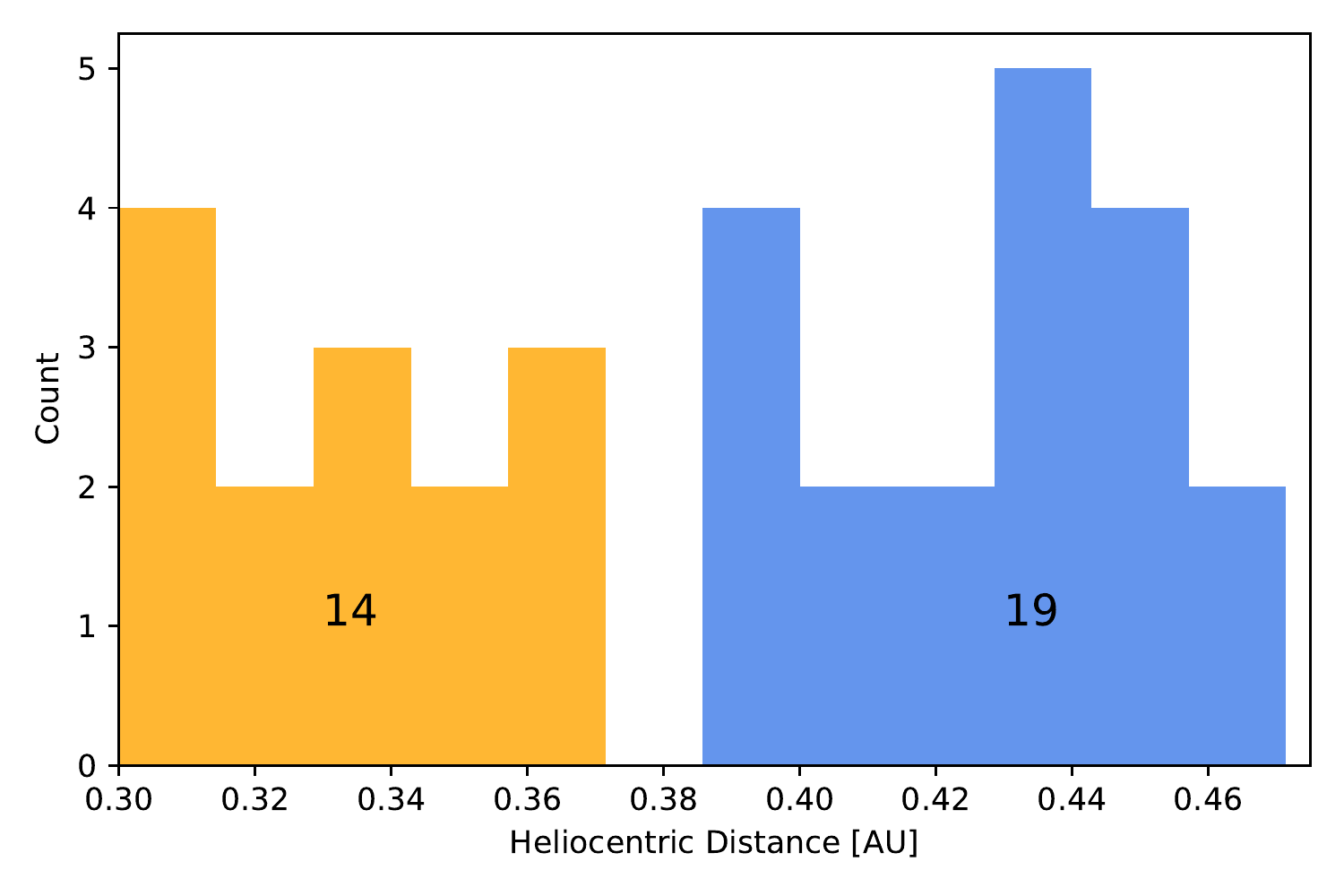}{0.45\textwidth}{(a) Distribution of Fd heliocentric distances.}
          \fig{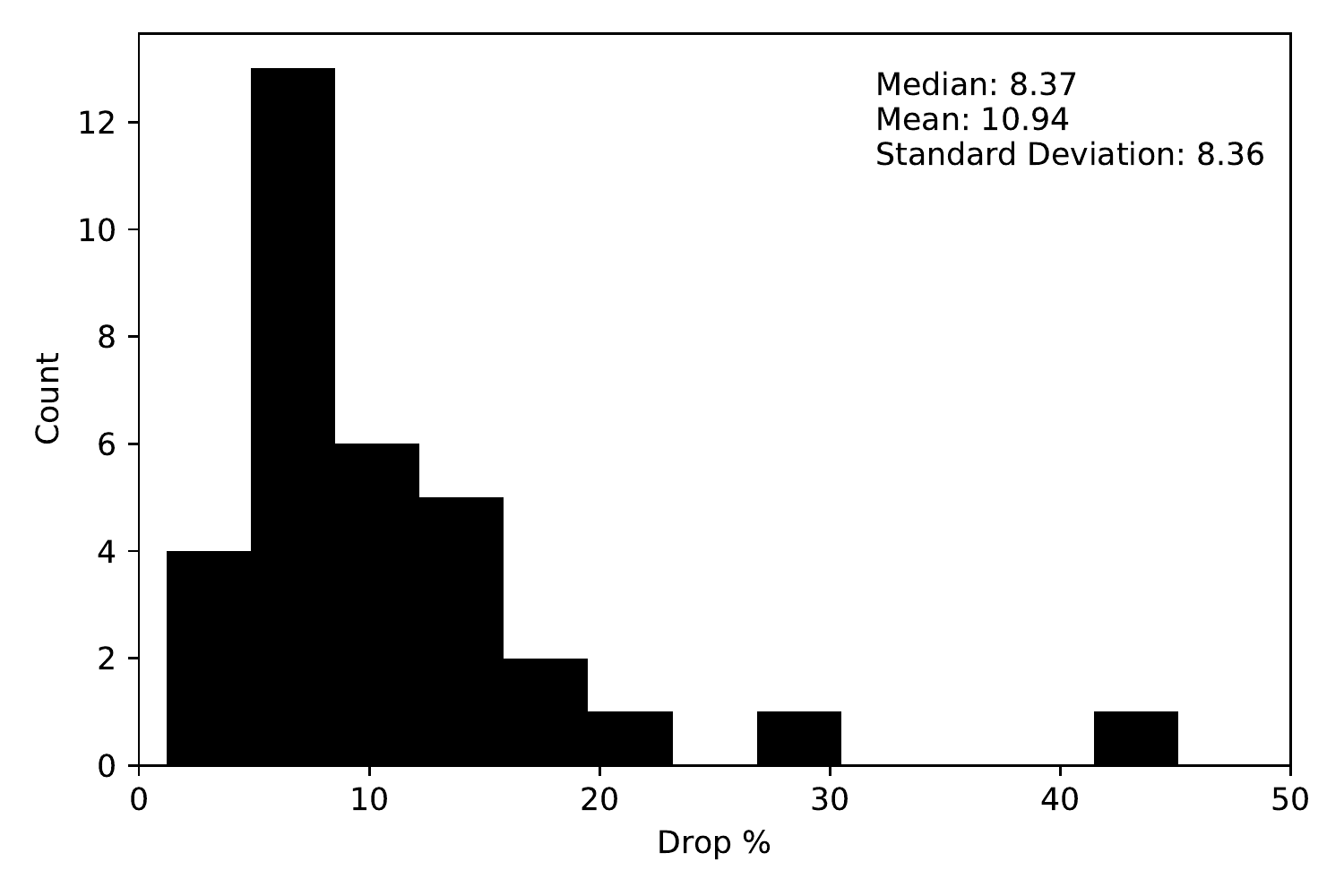}{0.45\textwidth}{(b) Distribution of Fd decrease percentages.}
          }
\gridline{\fig{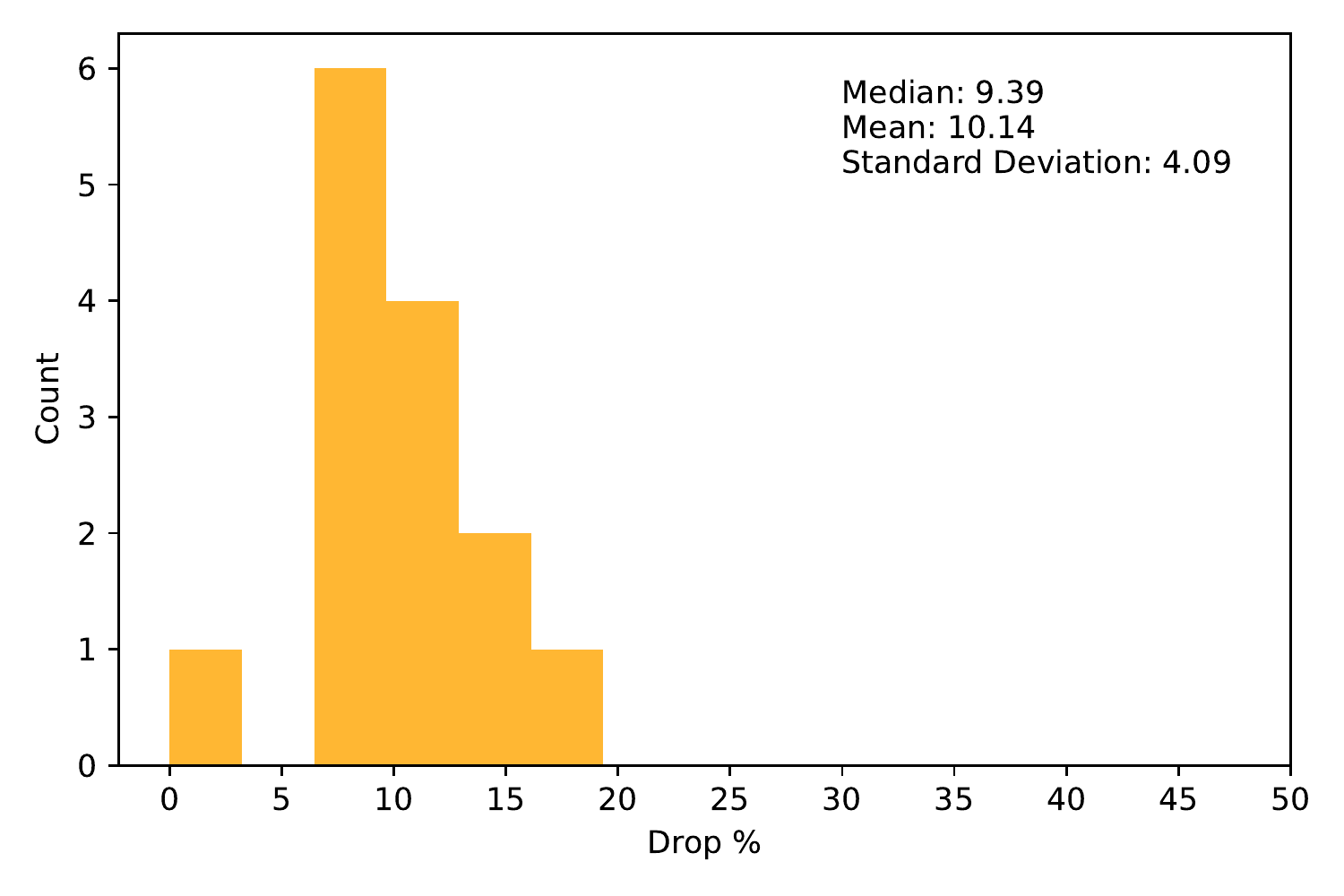}{0.45\textwidth}{(c) Distribution of Fd decrease percentages between 0.31 and 0.38~AU.}
          \fig{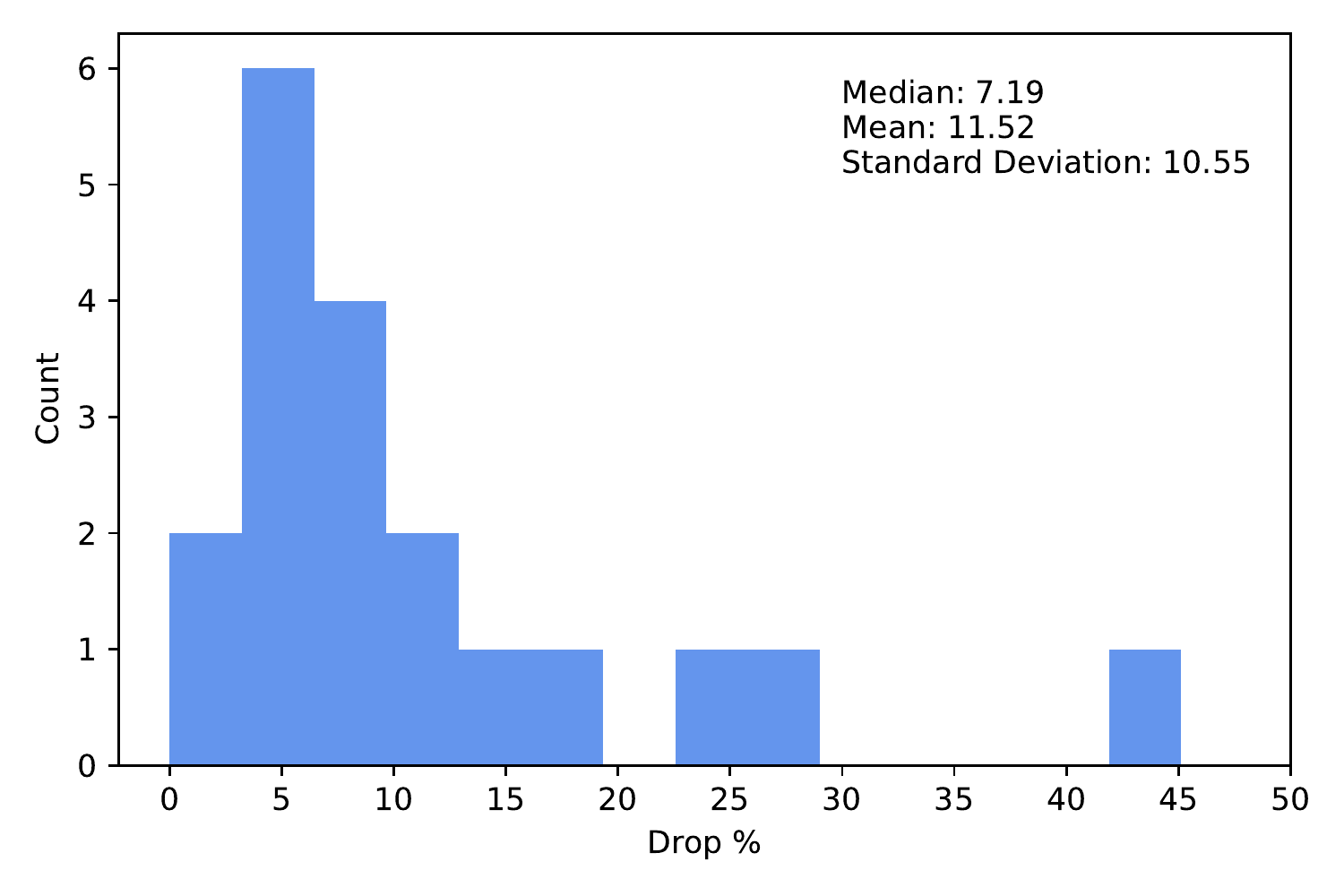}{0.45\textwidth}{(d) Distribution of Fd decrease percentages between 0.38 and 0.47~AU.}
          }
\gridline{\fig{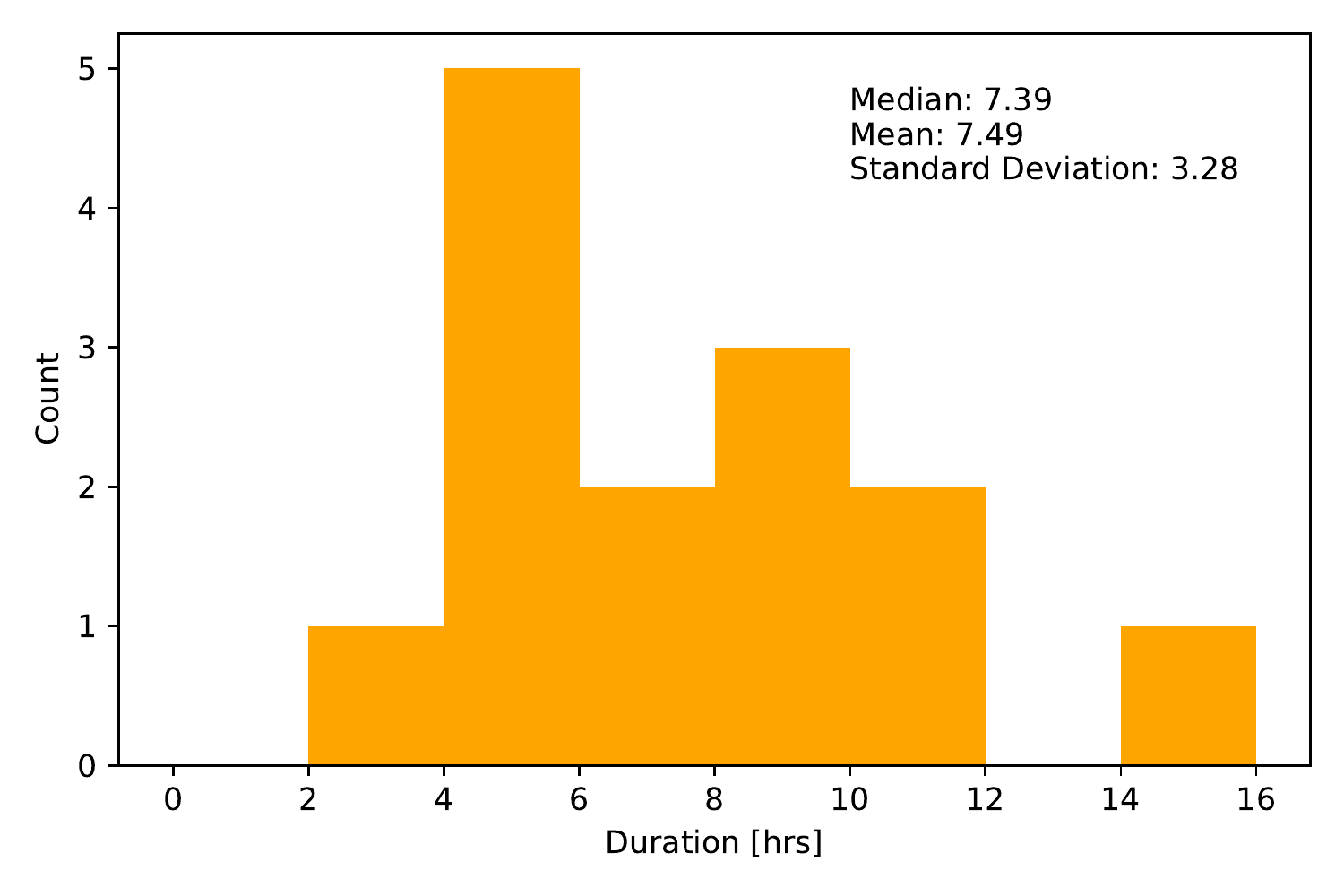}{0.45\textwidth}{(e) Distribution of Fd duration (time between onset and minimum) between 0.31 and 0.38~AU.}
          \fig{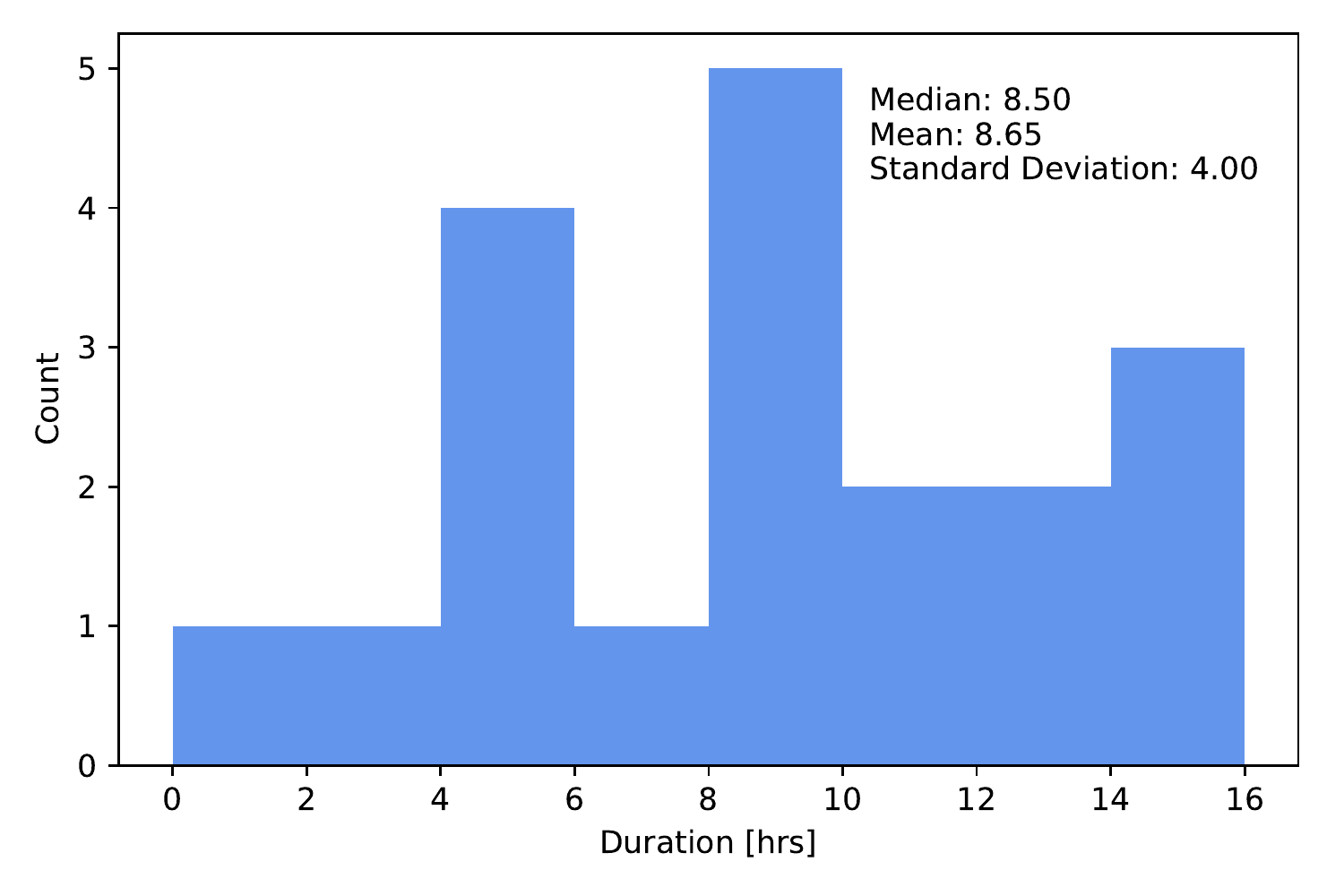}{0.45\textwidth}{(f) Distribution of Fd duration (time between onset and minimum) between 0.38 and 0.47~AU.}
          }
\caption{Histograms displaying the distribution of different cataloged Fd properties: heliocentric distance, percentage decrease and duration (time between onset and minimum). The median, mean and standard deviations are presented in the top right-hand corner of each panel, where appropriate. Black histograms represent all events identified in the Fd catalog, whereas orange and blue represent events observed at heliocentric distances less than and greater than 0.38~AU.}
\label{fig:fd_properties}
\end{figure*}

\begin{figure*}[t!]
\gridline{\fig{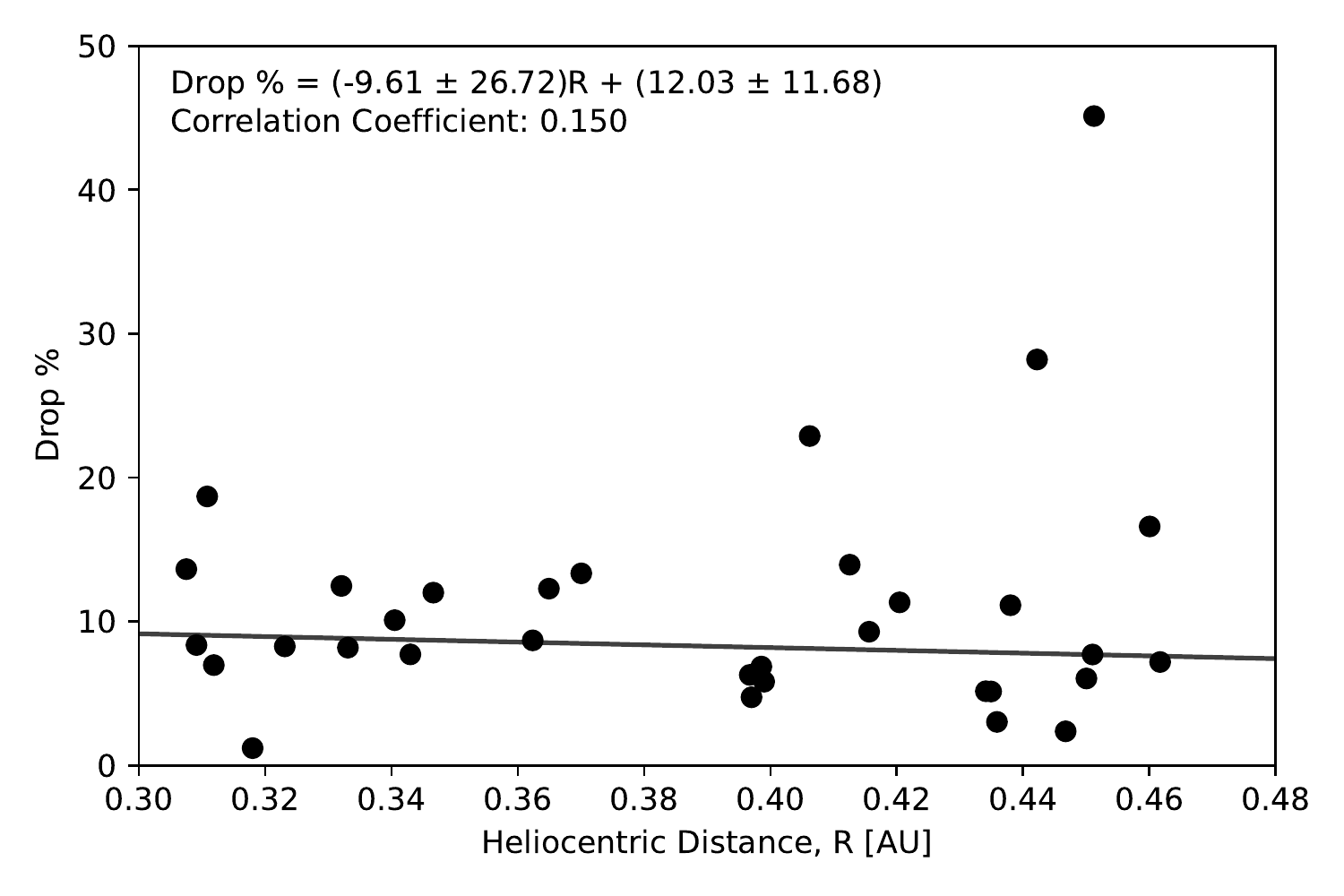}{0.45\textwidth}{(a) Fd percentage decrease with increasing heliocentric distance.}
          }
\gridline{\fig{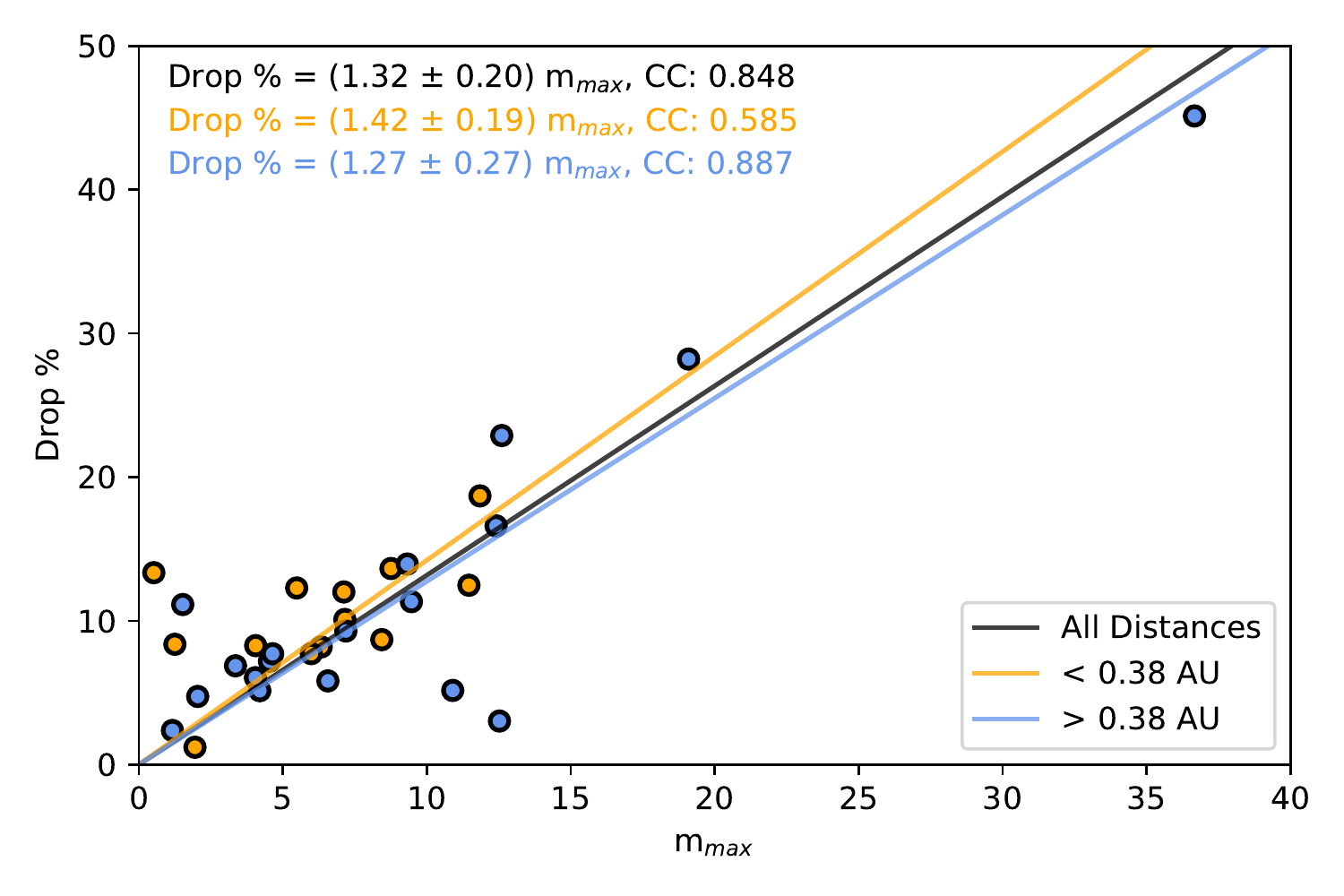}{0.45\textwidth}{(b) Fd percentage decrease against maximum hourly slope decrease.}
        \fig{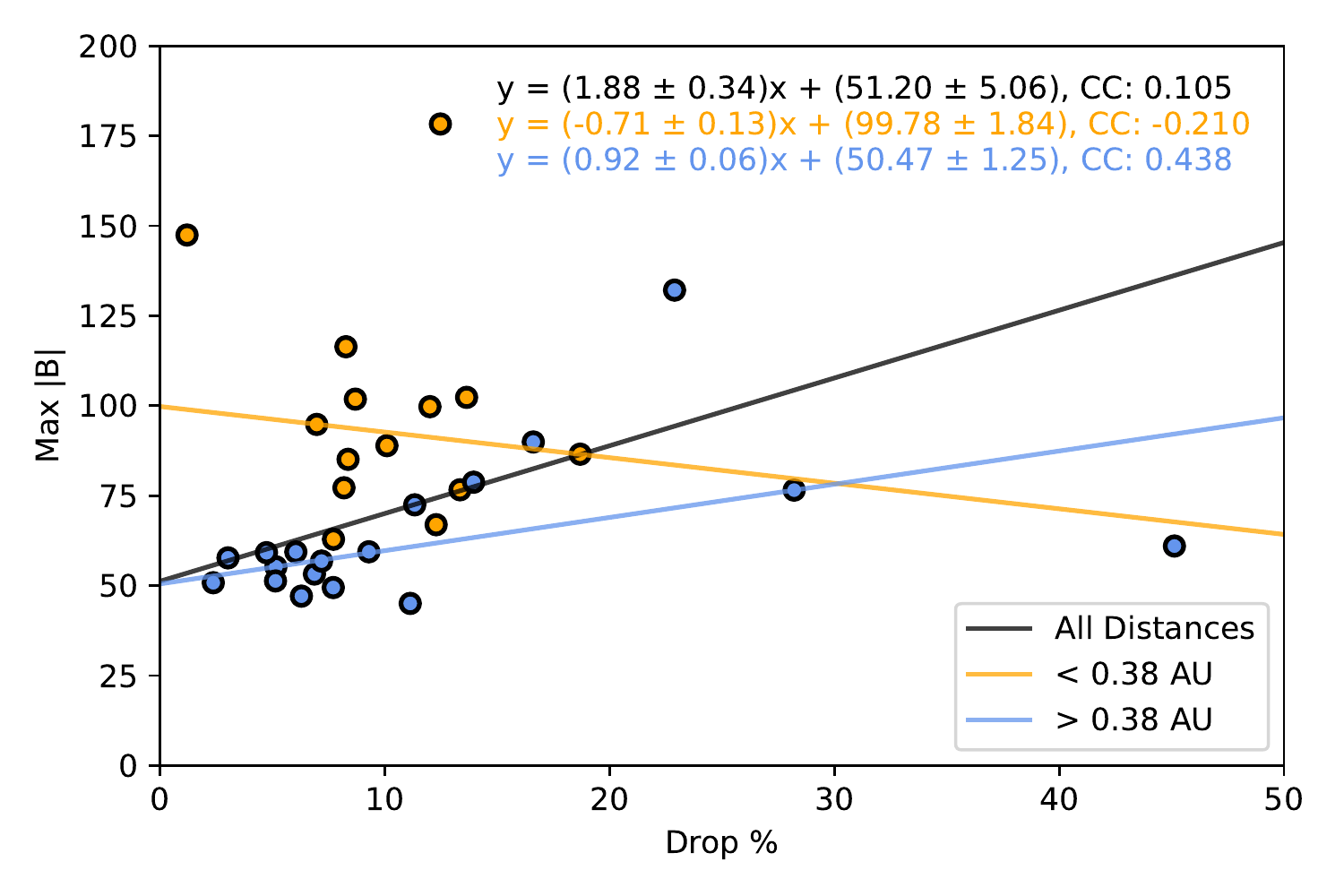}{0.45\textwidth}{(c) Maximum magnetic field observed within ICME associated with Fd against Fd percentage decrease.}}
\caption{Relationships between different properties, fit with a linear regression model. The equation of the line of best fit and associated uncertainties (the standard deviations), and the Pearson correlation coefficient are given in the top right hand corner for each figure.}
\label{fig:property_relationships}
\end{figure*}

Figures \ref{fig:fd_properties} and \ref{fig:property_relationships} present summary plots of the distributions and relationships between key parameters of the Fds identified, respectively. Figure \ref{fig:fd_properties}a presents the distribution of heliocentric distances at which events were identified. The eccentricity of Mercury's orbit covers heliocentric distances between 0.31 and 0.47~AU, and therefore, some Fd event properties may vary with increasing heliocentric distance. We assess the distribution of event distances and identify a split in the population around 0.38~AU: 14 events are identified between 0.31 and 0.38~AU, and 19 events between 0.38 and 0.47~AU. 

Figure \ref{fig:fd_properties}b presents the distribution of the percentage decrease in GCR count rate (drop percentage) for each Fd event identified. The mean percentage drop in GCR count value between onset and minimum was calculated to be $10.94 \pm 8.36$\%, with a median value of 8.37\%. Using the heliocentric distance split identified in Figure \ref{fig:fd_properties}a, we also investigate the distributions of drop percentage at closer heliocentric distances (0.31 to 0.38~AU; Figure \ref{fig:fd_properties}c) and further heliocentric distances (0.38 to 0.47~AU; Figure \ref{fig:fd_properties}d). We find the mean drop percentage at closer heliocentric distances to be $10.14 \pm 4.09$\%, and $11.52 \pm 10.55$\% at further heliocentric distances. This is contrary to what is expected with increasing heliocentric distances; one would expect the drop percentage to decrease with increasing heliocentric distance. However, the larger mean at further distances in this case is skewed by three events with drop percentages greater than 20\%, hence the large standard deviation calculated. Here, the median values give a better measure of most common expected drop percentage value in each distance range: we find the median drop percentages at closer and further distances to be 9.39\% and 7.19\%, respectively. The relationship between drop percentage and heliocentric distance is explored further in Figure \ref{fig:property_relationships}a. Using a least-squares fitting optimization model (scipy.optimize.least$\_$squares in Python) with a loss function of soft$\_$l1, we perform a robust fit to the data. We find a slight negative relationship, with a decrease in drop percentage with increasing heliocentric distance, as expected. However, the two parameters are very weakly correlated, with a Pearson's correlation coefficient of only 0.15.

Similar distributions were obtained to assess the mean duration of the Fds (time from onset to minimum). We find that the mean duration of the Fds is $8.16 \pm 3.70$~hrs, where the uncertainty given is the standard deviation. Using the same distance ranges as above, we calculate a mean duration of $7.49 \pm 3.28$~hrs at closer distances (Figure \ref{fig:fd_properties}e) and $8.65 \pm 4.00$~hrs at further distances (Figure \ref{fig:fd_properties}f), suggesting a slight increase in duration with increasing heliocentric distance.

We investigate the different slopes fitted for the two-step events identified. The 1st:2nd slope ratio indicates which of the slopes was steepest for each Fd event. We find that for 13 events (57\%), the 2nd slope was steeper than the 1st; the remaining 10 events (43\%) had steeper 1st slopes in comparison to the 2nd slope.

Another commonly used measure to assess the rate of decrease in GCR count rate during the Fd is known as the maximum decrease rate, $m_{max}$ \citep{belov2009forbush, abunin2012forbush, papaioannou2019forbush, vonforstner2020comparing}. To calculate $m_{max}$, we use the 24-hr rolling mean data to find the difference in GCR count rate between each datapoint, y(t$_i$), and the datapoint one hour earlier, y(t$_{i-1hour}$), for each event. The maximum of these values is then selected, normalized by the preceding background GCR count rate, and converted to a percentage value for comparison across all events. 

Figure \ref{fig:property_relationships}b presents the relationship between the percentage drop and $m_{max}$ for each event. Using the same least-squares fitting optimization model described above, we perform a robust fit to the data. We find a positive correlation, with a strong Pearson correlation coefficient of 0.85. The linear regression yields a gradient of $1.32 \pm 0.20$ hr. Splitting the dataset by distance around 0.38~AU, we find a gradient of $1.42 \pm 0.19$~hr for events between 0.31 and 0.38~AU, and $1.27 \pm 0.27$~hr between 0.38 and 0.47~AU. Previous studies comparing results at Earth and Mars have found that the gradient increases with increasing heliocentric distance \citep[][]{vonforstner2020comparing}, however, over the smaller distance split here, this trend is not observed. However, the gradients calculated are very similar within uncertainties, and therefore we select the gradient from fitting all events to compare with values calculated at Earth and Mars in Section \ref{sec:discussion}. 

Finally, Figure \ref{fig:property_relationships}c presents the relationship between the Fd drop percentage and the maximum magnetic field strength observed within the ICME. Using the same least-squares fitting optimization model as above, we perform a robust fit to the data where outliers have less influence on the model. We find the parameters are very weakly correlated with a Pearson correlation coefficient of 0.105, and a gradient of 1.88$\pm$0.34 between the two parameters. Splitting the dataset by distance emphasizes the weakness of the correlation between the parameters: for events between 0.31 and 0.38~AU, we find a weak negative correlation coefficient of -0.210, and for events between 0.38 and 0.47~AU, we find a weak positive correlation coefficient of 0.438.

\section{Superposed Epoch Analysis}

To determine the general profile of Fds observed by MESSENGER, we use the superposed epoch analysis \citep[SEA;][]{chree1913sea} technique. Although there are different variations of the technique, it essentially works by normalizing the time series of different events, allowing events to be superposed and an average profile to be calculated. The simplest version of SEA aligns time series at one common feature or boundary (e.g. a shock front or discontinuity). However, in the case of structures with multiple well-defined boundaries such as ICMEs (i.e. ICME start, ME start, ICME end) and Fds (i.e. Fd onset and Fd minimum), the time series can be normalized and scaled for each substructure. Such multi-bound SEAs have previously been performed to determine the average magnetic field and plasma profiles of corotating interaction regions \citep[CIRS;][using a two-bound SEA]{yermolaev2015seacir} and ICMEs \citep[][using a three-bound SEA]{janvier2019generic, regnault2020sea_ace}.

In Section \ref{subsec:sea_magnetic} we use the magnetic boundaries of the ICME when conducting the SEA, and in Section \ref{subsec:sea_fd} we use the Fd boundaries defined in this study. Due to the smaller sample size of events identified in this study, the events used in each SEA are a mix of quality ratings, with no one rating in the majority. Thus, we are unable to produce different SEA profiles based on each quality rating. To account for the data gaps present in lower quality events, we linearly fill values so as a representative value of GCR count rate is given for that particular part of the Fd profile, allowing for the largest possible sample size to be used when performing the SEAs.

\subsection{Superposed Epoch Analysis using Magnetic ICME Boundaries} \label{subsec:sea_magnetic}

Previous studies have used multi-bound SEA to relate the substructure of ICMEs to that of Fds \citep{masias2016superposed, janvier2021twostep}. Both \citet{masias2016superposed} and \citet{janvier2021twostep} used the magnetic boundaries of ICMEs to perform a three-bound SEA. This allowed \citet{masias2016superposed} to determine that for faster ICMEs, the average profile showed that a second minima in the GCR flux was observed within the ME region, and \citet{janvier2021twostep} found that ICMEs without sheath regions produced a more symmetrical average Fd profile, with the minimum GCR flux located closer to the center of the ICME.

To investigate the Fd profile in relation to the ICME substructure, we use only events where three ICME boundary times have been defined: ICME start, ME start, and ICME end. We further refine the events analyzed, removing events with interacting ICMEs and ICMEs in very close succession. A total of 25 events were used in the analysis. Of these 25 events, 19 (76\%) were identified in the database as having at least a two-step structure.

We define four regions of interest: pre-ICME, sheath, ME, and post-ICME. We perform the SEA time series normalization similarly to that of \citet{masias2016superposed}, \citet{regnault2020sea_ace}, and \citet{janvier2021twostep}: the normalized time unit is defined as the duration of the sheath (between ICME start and ME start). Similarly, we normalize the time series data within the ME region (between ME start and ICME end), and scale this to be three times the length of the sheath duration. We justify this scaling length based on the results of \citet{janvier2019generic}, which determined the ratio of average sheath to ME duration at MESSENGER as 1:3. The same normalization and scaling using the ME start and ICME end boundaries is also applied to the pre- and post-ICME regions to capture the behavior of the pre-ICME background GCR count levels and the recovery of the GCR count rate, respectively. 

With the time series normalized, the GCR count rate data for each event is averaged into bins of 0.1 times the unit normalized time e.g. 10 bins within the sheath, and 30 within the ME region. In the case of data gaps, we linearly fill values so as a representative value of GCR count rate is given for that particular part of the Fd profile, necessary when calculating the overall average profiles. It is important to note that the background GCR count rate can vary for each event, and therefore each binned value is also normalized by the mean preceding background value calculated over the 24 hours prior to the onset of the Fd. The GCR count rate is therefore presented as a percentage of the background count rate.

\begin{figure*}[t!]
\centering
\includegraphics[width = \textwidth]{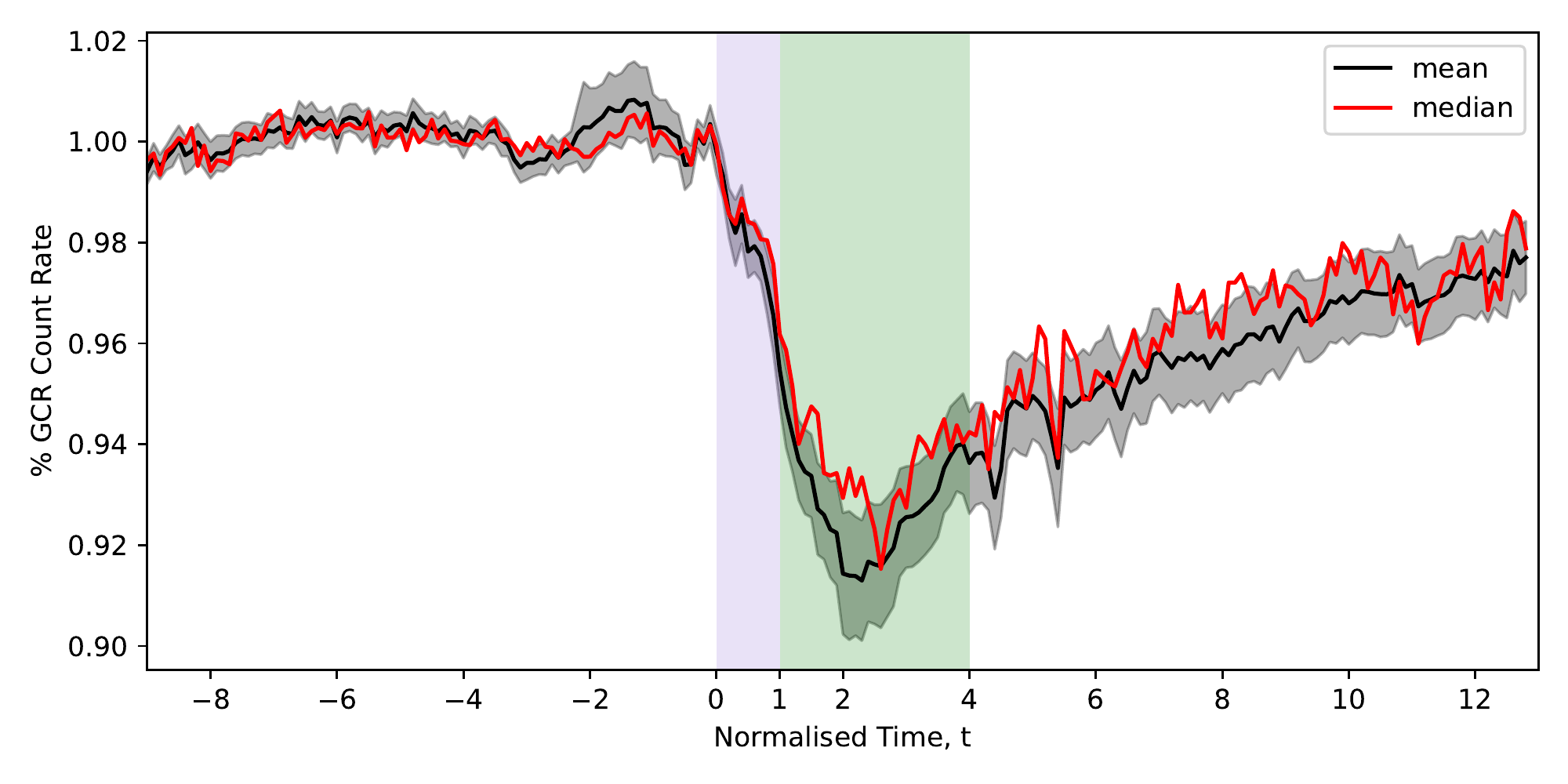}
\caption{Superposed epoch analysis of the GCR count rate associated with the passage of an ICME at MESSENGER. Similarly to Figures \ref{fig:gcr_data} and \ref{fig:two_examples}, the sheath region of the ICME is shaded in purple (0~$<$~t~$<$~1), and the magnetic ejecta region is shaded in green (1~$<$~t~$<$~4). The mean profile is given by the black line with associated standard errors shaded in grey, and the median profile by the red line.}
\label{fig:sea_mag_bounds}
\end{figure*}

The mean and median values for each bin across all events were then calculated to build the general profile of an Fd at MESSENGER. Figure \ref{fig:sea_mag_bounds} presents the results of the SEA. The sheath region extends between 0 and 1, and the ME region between 1 and 4, shaded similarly to Figures \ref{fig:gcr_data} and \ref{fig:two_examples} in green and purple, respectively. The mean profile is presented by the black line with the standard errors of the mean shaded in grey, and the median profile by the red line. For a normal distribution of values within each normalized time bin, the mean and median would be equal. However, in our study, with only 25 events used in this case, the distributions of values within each bin are not symmetric nor normal. We therefore present both the mean and median as two measures to better describe the distribution \citep[see discussion in][]{regnault2020sea_ace}.

We note that the time normalizations and chosen scaling for the sheath and ME regions are independent of each other, and therefore any comparison of slope gradients within these regions should be made with caution. However, both the mean and median profiles of Figure \ref{fig:sea_mag_bounds} display a steep initial drop almost coincident with the start of the sheath region, before a slight plateau and a second drop towards the end of the sheath region. Thus the SEA profiles display a two-step structure, strongly linked with the ICME magnetic substructure. 

% FIGURE 4: 

\subsection{Superposed Epoch Analysis using GCR Fd Boundaries} \label{subsec:sea_fd}

\begin{figure*}[t!]
\gridline{\fig{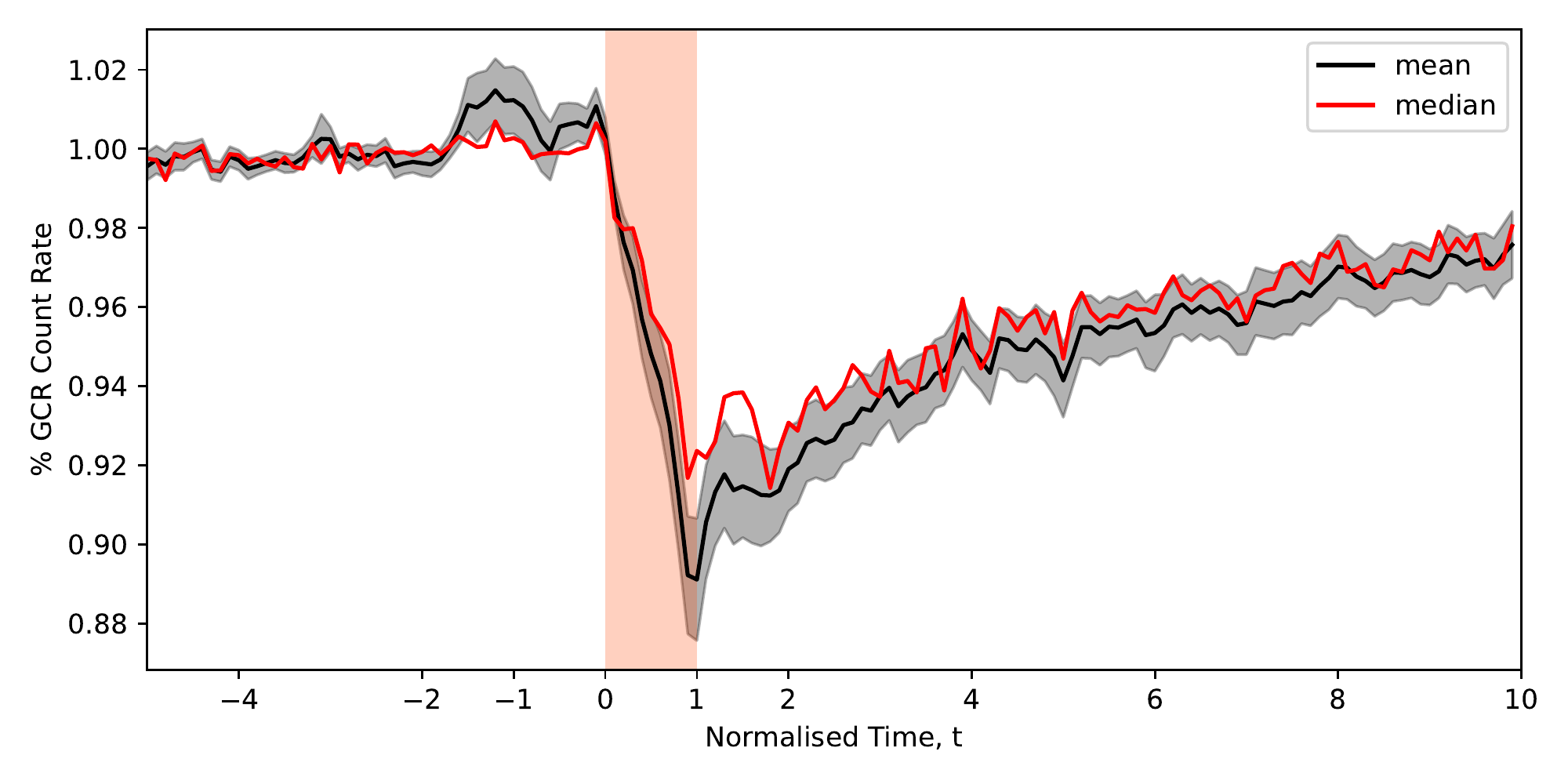}{0.5\textwidth}{(a) Two bound SEA, normalised between Fd onset and minimum.}
          \fig{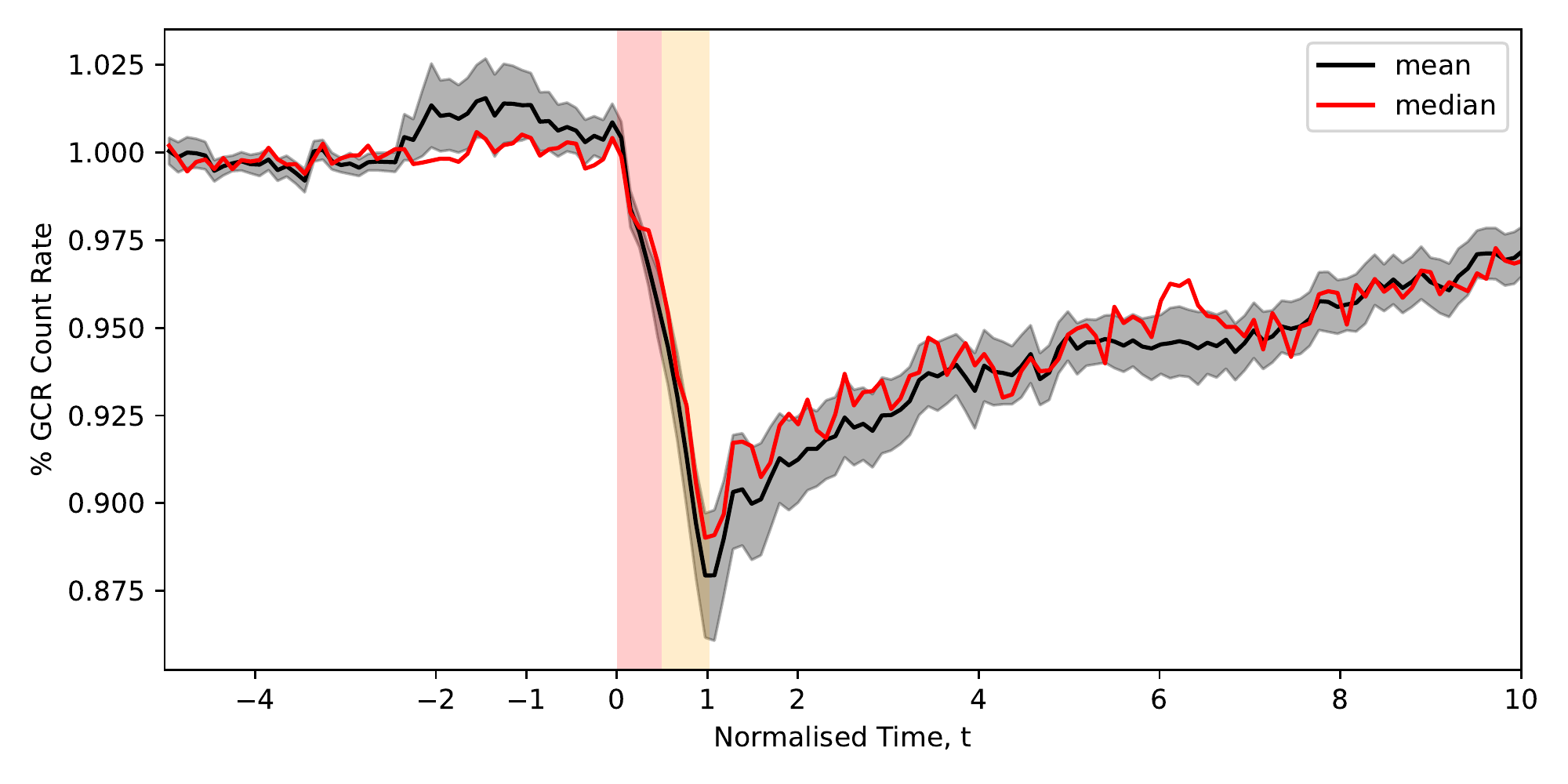}{0.5\textwidth}{(b) Three bound SEA, normalised between Fd onset and second drop time, and second drop time and minimum.}
          }
\caption{Superposed epoch analysis of the GCR count rate using the Fd boundaries defined in Table \ref{tab:database}. }
\label{fig:sea_fd_bounds}
\end{figure*}

To further investigate this two-step Fd structure, we also perform a SEA using the Fd boundaries defined in Table \ref{tab:database}. Figure \ref{fig:sea_fd_bounds}a presents a two-bound SEA, normalizing between the Fd onset and minimum times (shaded in orange), and Figure \ref{fig:sea_fd_bounds}b presents a three-bound SEA normalizing between the Fd onset and second drop times (shaded in red), and the second drop time and Fd minimum (shaded in yellow). Similarly to Figure \ref{fig:sea_mag_bounds}, both the mean profile with associated standard errors (black) and the median profile (red) are presented. 

The mean and median SEA profiles of Figure \ref{fig:sea_fd_bounds}a do not display a clear two-step structure; the mean slope displays a relatively steady descent between onset (t=0) and the minimum (t=1), whereas the median profile displays three separate steps between onset and the minimum, however, these steps are likely due to the more variable median profile in comparison to the mean profile. For this analysis, we included every Fd listed, excluding those where ICMEs were in close succession or interacting, contributing to the same Fd; a total of 27 events, of which 21 (84\%) were identified as having at least a two-step structure. By including both one-step and two-step Fds, it is therefore unsurprising that a clear average two-step profile is not produced despite 84\% of the Fds used in this analysis having greater than one-step profiles. 

The two-step profile is also likely not present in Figure \ref{fig:sea_fd_bounds}a due to the different times at which the second drop occurs. To investigate this further, we perform a three-bound SEA using only events with at least a two-step structure (total of 21 events), normalizing between the Fd onset and second drop time, and second drop time and Fd minimum. The average profiles of the SEA are presented by Figure \ref{fig:sea_fd_bounds}b. The different normalizations of the three-bound SEA have been scaled to the average duration ratio i.e. onset to second drop: second drop to minimum. This was found to be approximately even: 1 to 1.03. Despite including only events with at least a two-step structure, this is not reflected clearly in either the mean or median SEA profiles. The median profile does display a possible two-step structure, with an initial drop, slight plateau, then drop again within the shaded red region, however, the single datapoint creating the plateau feature in the SEA median profile lies outside of the shaded standard error region, and similarly to Figure \ref{fig:sea_fd_bounds}a, is likely to be due to the more variable median profile. Overall, using the defined Fd boundaries, the two-step average profile is not as clear as that obtained using the magnetic ICME substructure boundaries.

% FIGURE 5:

\section{Discussion and Conclusion} \label{sec:discussion}

In this study, we characterize the structure of Fds at Mercury using the GCR count rate data onboard MESSENGER, building an average profile using a superposed epoch analysis, and attempt to relate the two-step Fd structure to the ICME substructure at MESSENGER. By using results from previous studies at Earth and Mars, here we also aim to address whether two-step Fds are more commonly observed closer to the Sun, and how Fd profiles and properties change with increasing heliocentric distance. 

Analyzing the GCR dataset, we found that 72.7\% (24/33) of Fds identified at MESSENGER (0.31--0.47~AU) had a possible two-step or more complex structure. Splitting the database by the same heliocentric distance used in Section \ref{sec:properties} (0.38~AU), we find similar numbers of two-step Fds at closer (0.31--0.38~AU) and further (0.38--0.47~AU) distances from the Sun: 10/14 (71.4\%) and 14/19 (73.7\%), respectively. At Earth, \citet{cane1996cosmic} studied 30 years of neutron monitor data between 1964 and 1994, identifying a total of 153 Fds with GCR flux decreases greater than 4\%, 86\% of which were caused by CMEs. Of these, 92 (70\%) were classified as two-step Fds. The percentage of two-step Fds they observed at Earth is therefore similar to the percentage we observe at MESSENGER. However, \citet{jordan2011revisiting} revisited the prevalence of two-step Fds at Earth using cosmic ray data from 14 neutron monitors and the International Gamma Ray Astrophysical Laboratory (INTEGRAL) mission where possible. They performed analysis of 233 ICMEs, 80 of which drove Fds, and found that only 13 (16.2\%) of these profiles were two-step Fds. This is a significantly lower percentage of two-steps than identified by \citet{cane1996cosmic}, and may arise from the stricter selection methods used by \citet{jordan2011revisiting} in identifying two-step events: \citet{jordan2011revisiting} analyzed only those ICMEs observed between February 1998 and December 2006 which had shocks and a sheath region longer than 4 hours, from which a two-step Fd could be identified if a step occurs within 2 hours of a shock or leading edge of the ICME and the same step was observed across the majority of four Bartol neutron monitors (McMurdo, Newark, Thule, and South Pole). Results obtained by \citet{jordan2011revisiting} would therefore suggest that two-step Fds may be more common at MESSENGER (72.7\%) closer to the Sun than at Earth, however, further studies at Earth are needed to resolve the discrepancy between the \citet{cane1996cosmic} and \citet{jordan2011revisiting} results, as well as larger statistical studies with a greater number of events identified near Mercury to determine whether two-step Fds are truly more common at lesser heliocentric distances.

Despite prevalence of two-step Fd profiles observed at MESSENGER, we find that performing a two-bound SEA of the GCR data (Figure \ref{fig:sea_fd_bounds}a) using the defined Fd boundaries (between onset and minimum) results in an average profile that does not display a clear two-step structure. This can be explained by the individual variation of each Fd, where the second drop time may occur in different places between onset and minimum, and the fact that the ratio between 1st and 2nd slope also varies with each event, affecting the average profile. To account for this, we also perform a three-bound SEA using the onset, second drop, and minimum boundaries for only those Fds identified as two-step profiles. However, the resulting average profile is not any clearer using this method (Figure \ref{fig:sea_fd_bounds}b), as the varying ratio between the 1st and 2nd slopes for each event still remains a factor affecting the average SEA profile. \citet{jordan2011revisiting} suggests that the traditional simple model of two-step Fds may need to be replaced, as the small-scale magnetic field structure of the ICMEs contributes to the variability of the Fd profiles. This variability and more complex structure of Fd profiles is reflected in our study with the inclusion of a complex step category where there may be more than two-steps. To investigate how the ICME magnetic substructure may be related to the structure of Fds, we perform another three-bound SEA using the ICME boundaries (ICME start, ME start, ICME end) for the ICME events causing these Fds defined by \citet{winslow2015interplanetary, winslow2017magnetosphere}. Using this methodology, we observe a strong link between the ICME substructures and GCR modulation, with a clear two-step average Fd profile observed (Figure \ref{fig:sea_mag_bounds}). When comparing the sheath and ME regions of the SEA in \ref{fig:sea_mag_bounds}, caution must be taken when comparing the slopes because of the different time normalizations and scaling. However, most of the structure related to the two-step profile occurs within the sheath region; we observe an average profile with a clear initial drop at the start of the sheath region, a plateau, and the onset of the second drop occurs towards the end of the sheath region. This result is in line with previous studies that suggest drops in GCR flux are due to diffusive barriers created by the shock and sheath, and later the discontinuity between the sheath and the leading edge of the ME \citep[e.g.][]{sanderson1990relation, wibberenz1997twostep}, linking the two-step Fd to the magnetic substructure of the ICME.  

To investigate how GCR modulation and thus the Fd profiles change as the ICME propagates and evolves, we also perform both the SEAs using the magnetic and Fd boundaries discussed above but over the two distance ranges used throughout (not shown). The average profiles produced are very similar to those found using every event in Figures \ref{fig:sea_mag_bounds} and \ref{fig:sea_fd_bounds}, respectively. Interestingly, the average profiles using either method are similar over both distance ranges; there is no difference to the resulting average two-step profile closer or further from the Sun over the distance range covered by MESSENGER. However, the main difference is the average percentage GCR count rate decrease at the minimum: we find this to decrease with increasing heliocentric distance. This result is shown by the histogram plots of Figures \ref{fig:fd_properties}c and d, where the median drop percentage decreases with increasing heliocentric distance from 9.39\% between 0.31 and 0.38~AU to 7.19\% between 0.38 and 0.47~AU. The mean values for each distance range were found to be 10.14\% and 11.52\%, respectively, where the increase with distance in this case is due to three larger Fd events skewing the mean between 0.38 and 0.47~AU. The overall mean drop percentage (every event included) was found to be 10.94\%. This result is directly comparable to the mean drop percentage found using a catalogue of ICME-associated Fds observed by MAVEN in orbit around Mars of 6.37\% \citep{guo2018marsfds}, suggesting that the Fds decrease in amplitude with increasing heliocentric distance. This is in agreement with the findings of previous studies that tracked the Fd signatures driven by a single ICME observed at MESSENGER, LRO, the SOPO Neutron Monitor and MSL \citep{winslow2018window}, as well as for another ICME observed at Mars, comet 67P, and
Saturn \citet{witasse2017interplanetary}, where percentage decreases in GCR flux were found to decrease with increasing heliocentric distance for both ICME cases studied.

Simple models of perpendicular diffusion across the ME suggest the maximum decrease in GCR flux of the Fd (percentage drop) decreases as $Bva^2$, where $B$ is the magnetic field strength of the ME, $v$ is the ICME speed, and $a$ is the radius of the flux rope; we therefore expect the Fd drop percentage to be less when driven by ICMEs with weaker magnetic field strengths, smaller radial widths, and slower propagation speeds. With only magnetic field data at MESSENGER, we focus on how the Fd drop percentage may be related to the ICME maximum magnetic field strength. Previous studies have suggested that the intensity of the magnetic field strength of an ICME is the necessary condition in driving a Fd \citep[e.g.][]{janvier2021twostep}, and \citet{belov2015galactic} found that all ICMEs with ME maximum magnetic field strengths $>$18~nT (and ICMEs with MC maximum magnetic field strengths $>$20~nT) at 1~AU drove Fds with maximum GCR percentage decreases within the ME region. In this study, all ICMEs associated with the identified Fds have maximum magnetic field strengths $>$18~nT (note this value was established at 1~AU, and therefore there may be a different minimum magnetic field strength required to drive Fds at Mercury); the minimum ICME B$_{max}$ observed in this study that drove a Fd was 45~nT. We further investigate whether there is a difference in B$_{max}$ of ICMEs that drove Fds and those that did not. Including only ICMEs with corresponding good quality GCR data (a total of 42), we found that those that drove Fds (33, 78\%) had a mean B$_{max}$ of 79.4$\pm$30.9~nT and a median of 74.5~nT, in comparison to those that did not drive a Fd with a mean of 67.0$\pm$14.7~nT and a median of 62.1~nT. This result shows that overall, that Fd-driving ICMEs were of higher maximum magnetic field strengths on average than those that did not drive Fds, however, the mean values are not significantly different with overlapping uncertainty ranges. We explore the relationship between magnetic field strength and GCR modulation further in Figure \ref{fig:fd_properties}f, by plotting the maximum magnetic field strength against percentage drop for each ICME. Performing a robust linear fit to the data, we find a positive correlation between the two parameters, however, this correlation is very weak with a Pearson correlation coefficient of only 0.11. We investigated whether the different distance ranges have any effect on the relationship, and found a similarly weak positive correlation for events between 0.38 and 0.47~AU (0.44), and a weak negative correlation for those between 0.31 and 0.38~AU (-0.21). 

To assess how the rate of decrease in GCR flux within the Fd changes with distance, we calculated the maximum decrease rate, $m_{max}$. We find that the maximum decrease rate has a strong positive correlation (0.85) with the maximum percentage decrease in GCR flux within the Fd for events in our database at MESSENGER, similarly to previous studies at Earth and Mars \citep{belov2009forbush, abunin2012forbush, papaioannou2019forbush, vonforstner2020comparing}. We compare the gradient of the fitting ($1.17 \pm 0.06$) at MESSENGER, to those calculated for ICME-driven Fds by \citet{vonforstner2020comparing} at Earth and Mars: $4.60 \pm 0.27$ and $6.80 \pm 0.82$, respectively. Using these results, we find that the gradient of the fitting ($\frac{\Delta drop \%}{\Delta m_{max}}$) has a positive linear relationship with increasing heliocentric distance ($r$) of $\frac{\Delta drop \%}{\Delta m_{max}} = (5.13\pm1.56) r -(0.83\pm1.65)$. \citet{vonforstner2020comparing} relates the values of $m_{max}$ calculated, to the size of the ICME: as an ICME expands during propagation, the duration of the Fd (from onset to minimum) increases, and thus $m_{max}$ decreases for an Fd with the same percentage decrease in GCR flux at the minimum. This has the implication that the gradient of the linear relationship between percentage drop and $m_{max}$ increases as the ICME expands. \citet{vonforstner2020comparing} also notes that the gradient of the linear relationship between percentage drop and $m_{max}$ is unaffected by the GCR energies and the amplitude of the the percentage drop value of the Fd (see Figure 6 of \citet{vonforstner2020comparing} for visual explanation.) The resulting gradient found in this study fits well with the trend found between Earth and Mars; the lower gradient at MESSENGER is in line with smaller duration ICMEs at lesser heliocentric distances that expand as they propagate further away from the Sun. 

In conclusion, we have identified 33 Fd events associated with ICMEs observed by MESSENGER between 0.31 and 0.47~AU. In this study, we characterized the structure of Fds at Mercury and used a superposed epoch analysis to relate the two-step Fd structure to the ICME substructure at MESSENGER, finding that the average Fd profiles resulting from the SEA link the larger two-step structure directly with the magnetic ICME substructure of the sheath and ME regions. By using results from previous studies at Earth and Mars, we also aimed to address whether two-step Fds are more commonly observed closer to the Sun; we found that two-step Fds are possibly more commonly observed closer to the Sun but this is not conclusive when comparing to the wide range of results of previous studies conducted at Earth. Finally, we investigated how Fd profiles and properties change with increasing heliocentric distance; we found that properties of the Fd profile such as the percentage decrease in GCR count at the minimum is larger on average at MESSENGER, and the relationship between the percentage decrease and maximum hourly decrease ($m_{max}$) fit with trends of increasing ICME size as the ICME propagates further from the Sun. This is the most recent and thorough investigation of Fds within the inner heliosphere since the Helios mission, however, further investigation into the smaller scale magnetic field structure and the mechanisms that drive either a one-, two- step, or more complex Fd profile at these closer heliocentric distances is necessary. Analysis of the GCR dataset at MESSENGER in this study contributes to the understanding of Fds closer to the Sun, and provides a good basis on which other missions such as Solar Orbiter, Parker Solar Probe, and BepiColombo may build upon in the future. 

\begin{acknowledgments}

We thank the MESSENGER Neutron Spectrometer instrument team for the distribution of their dataset that made this study possible. This research was supported by NASA grants 80NSSC19K0914 (E.E.D. and R.M.W.) and NNX16AI98G (D.J.L.). E.E.D. would also like to thank H.J. Davies for his expertise in filtering the GCR dataset. The MESSENGER Fd database produced by this study is available at \url{https://doi.org/10.6084/m9.figshare.20477265}. 
 
\end{acknowledgments}

\vspace{5mm}
\software{spiceypy \citep{annex2020spiceypy}, scipy \citep{virtanen2020scipy}, scikit-learn \citep{scikit-learn}}

\appendix

\begin{figure*}
\centering
\includegraphics[width = \textwidth]{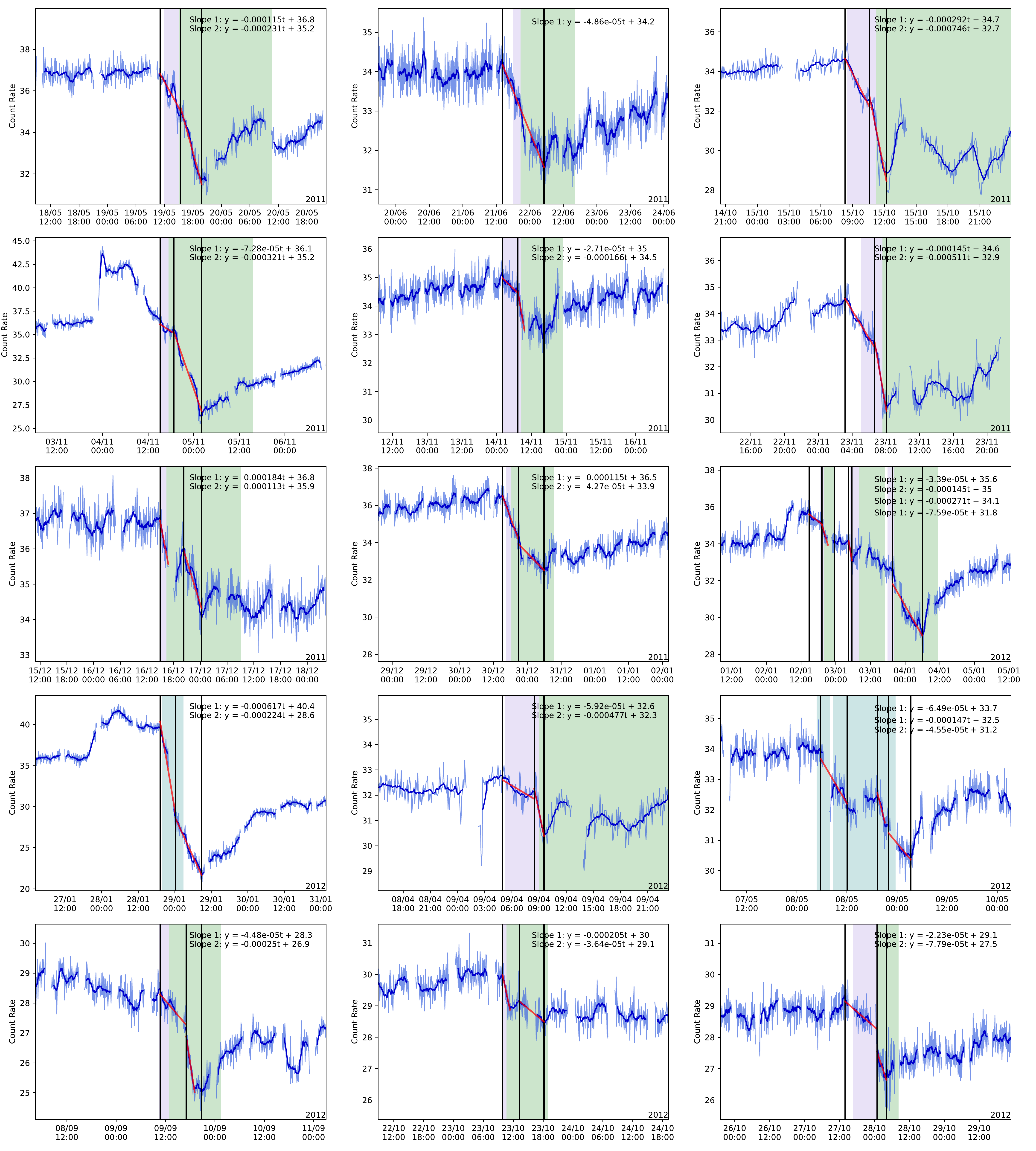}
\caption{Events 1--18. The ICME sheath and ME region are shaded in purple and green respectively, and vertical black lines mark the onset, second drop (where present) and minimum times in chronological order. The results of the linear regression fitting are overlaid in red and equations of the slopes presented in the top right-hand corner of each plot.}
\label{fig:appendix1}
\end{figure*}

\begin{figure*}
\centering
\includegraphics[width = \textwidth]{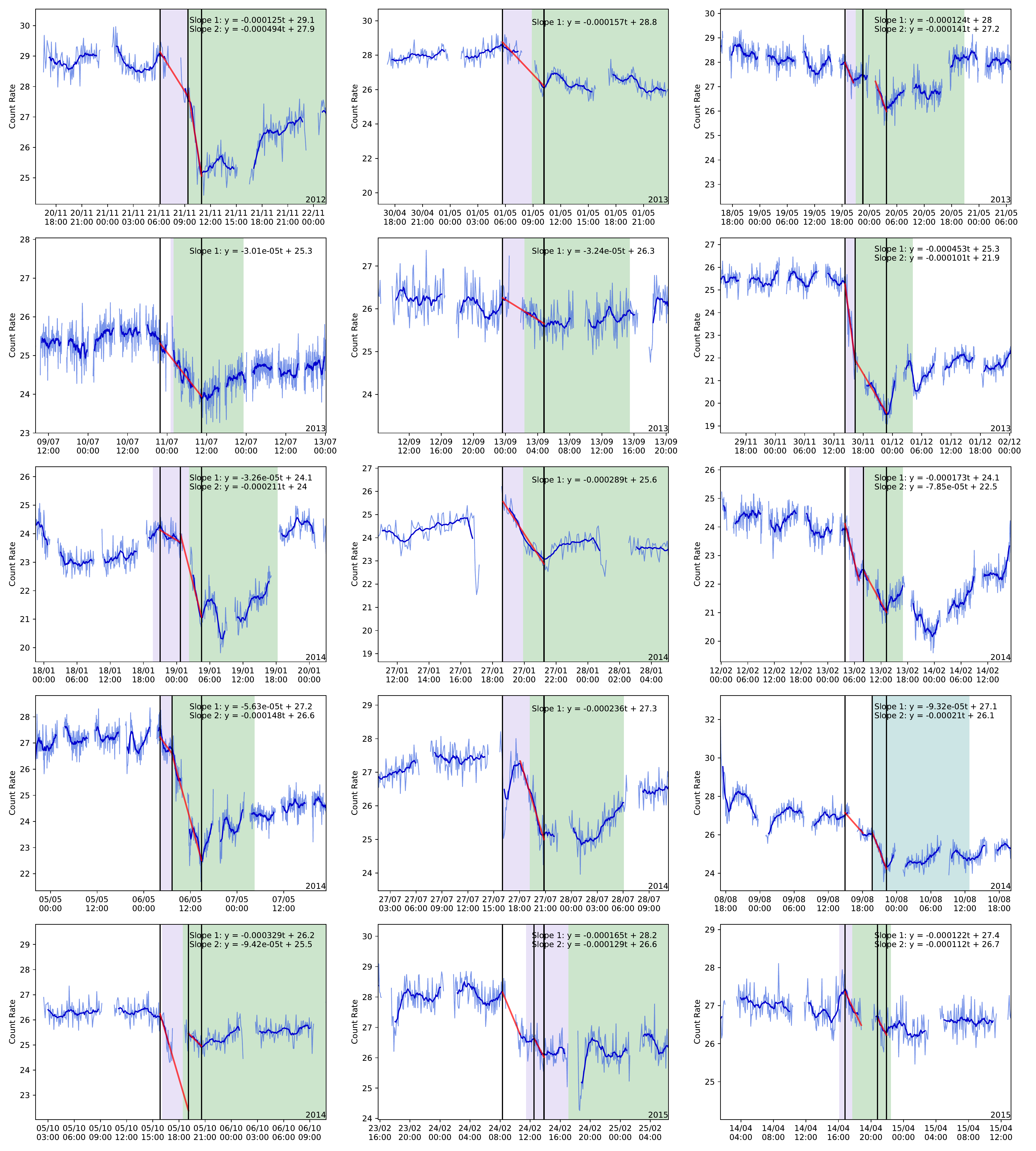}
\caption{Events 19--33. Presented in the same format as Figure \ref{fig:appendix1}.}
\label{fig:appendix2}
\end{figure*}

%\section{}

\bibliography{bibliography}{}
\bibliographystyle{aasjournal}


\end{document}